\documentclass[doublecol]{epl2}

\title{
    Non-equilibrium vortex annealing of structural disorder
    in Berezinskii-Kosterlitz-Thouless dynamics of two-dimensional XY-model
}

\shorttitle{
    Non-equilibrium structural disorder vortex annealing in BKT dynamics
} 

\author{
Popov I.S.\inst{1} \and 
Popova A.P.\inst{1} \and 
Prudnikov P.V.\inst{1}}
\shortauthor{Popov I.S. \etal}

\institute{                    
  \inst{1} Department of Theoretical Physics, Omsk State University --
  Omsk 644077, Russia
}


\newcommand{\mcs} {\rm MCS/s}
\newcommand{\TBKT} {T_{\rm BKT}}
\newcommand{\TBKTp} {T_{\rm BKT}(p)}

\graphicspath{{figs/}}

\abstract{
The non-equilibrium annealing of structural disorder in a two-dimensional 
XY-model leads to coarsening of defects clusters in a cores of spin vortices.
We revealed the effect of ``inertial"{} growth of the clusters in coarsening dynamic regime.
The calculated transverse stiffness $\rho(p,T)$ of the system in the high-temperature phase $T>T_{\rm BKT}(p)$ becomes negative and has described by a power low $\rho(p,T) \sim T^{-\kappa}$ with temperature independent exponent $\kappa=\kappa(p)$.
The dynamical scaling lead to the dynamic dependence of the correlation length $\xi \sim (t / \ln^{q} (t/t_{0}))^{1/z}$ which can be explained by a shift of  
spin vortices friction constant $\gamma$ induced by annealed disorder.
}

\begin{document}

\maketitle


Investigations of critical behavior of 
systems with continuous symmetry of the order parameter attract a 
lot of attention and represent considerable fundamental and practical interest~\cite{MaS}.
Two-dimensional systems with continuous symmetry occupy a special 
place among low-dimensional systems. It is well known that the long-range 
order is broken in these systems at any finite temperature. 
However, the two-dimensional XY-model is characterized by realization of topological 
Berezinskii-Kosterlitz-Thouless (BKT) phase transition at temperature $\TBKT$%
~\cite{Berezinskii_1,Berezinskii_2,Berezinskii_book,Kosterlitz_Thouless,Kosterlitz}.

\begin{figure}[ht!]
	\centering
	\includegraphics[width=0.23\textwidth]{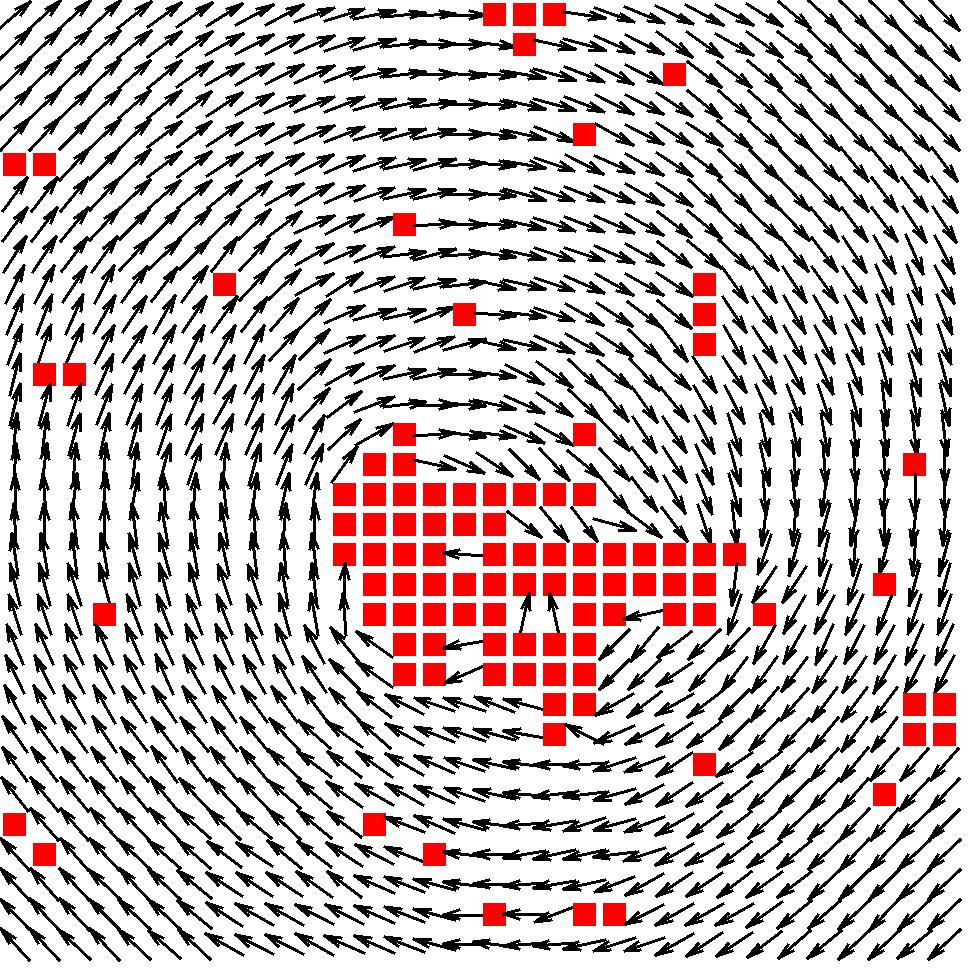}\hspace{1mm}
	\includegraphics[width=0.23\textwidth]{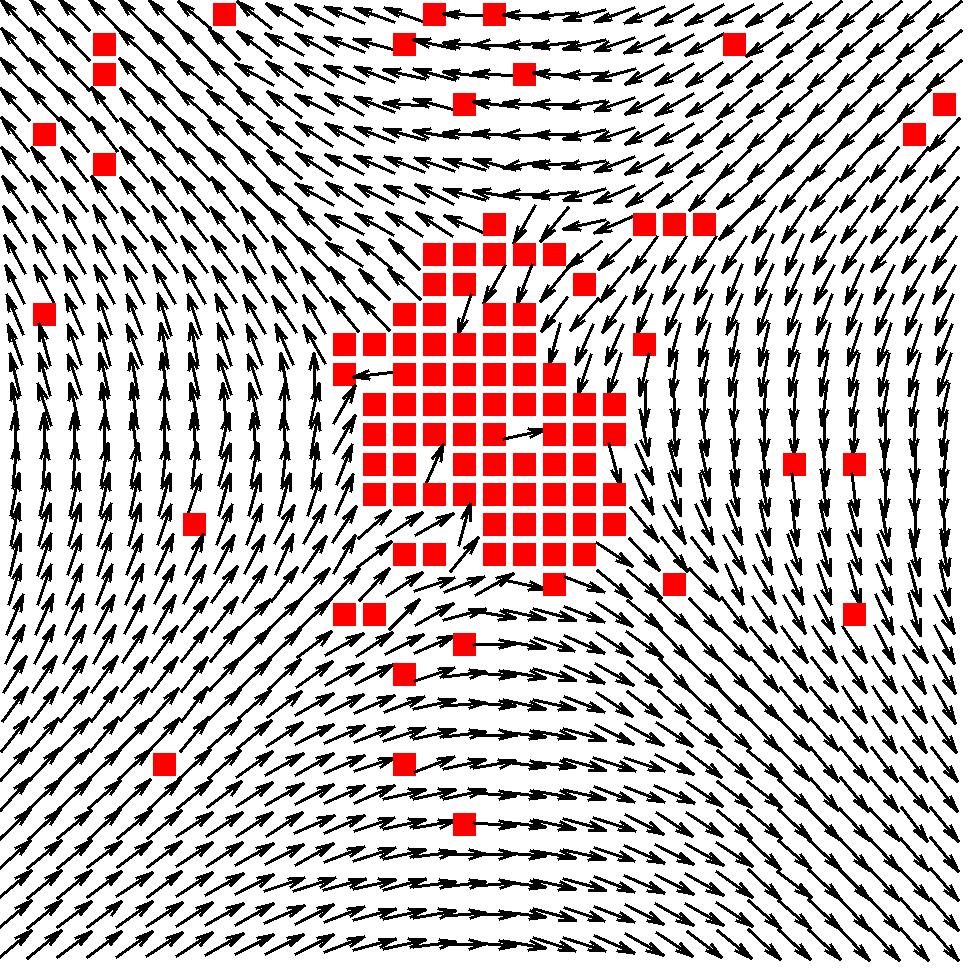}
	\caption{\label{Snapshot_lc}
		Typical snapshot configurations of large clusters of defects growths
		by interaction of structural disorder with vortex (left) and antivortex (right) 
		in the system with the linear size $L = 32$ 
		and the spin concentration $p = 0.9$.
		Arrows demonstrate spins, red squares -- defects.
	}
\end{figure}
\begin{figure*}[t!]
	\centering
	\includegraphics[width=0.23\textwidth]{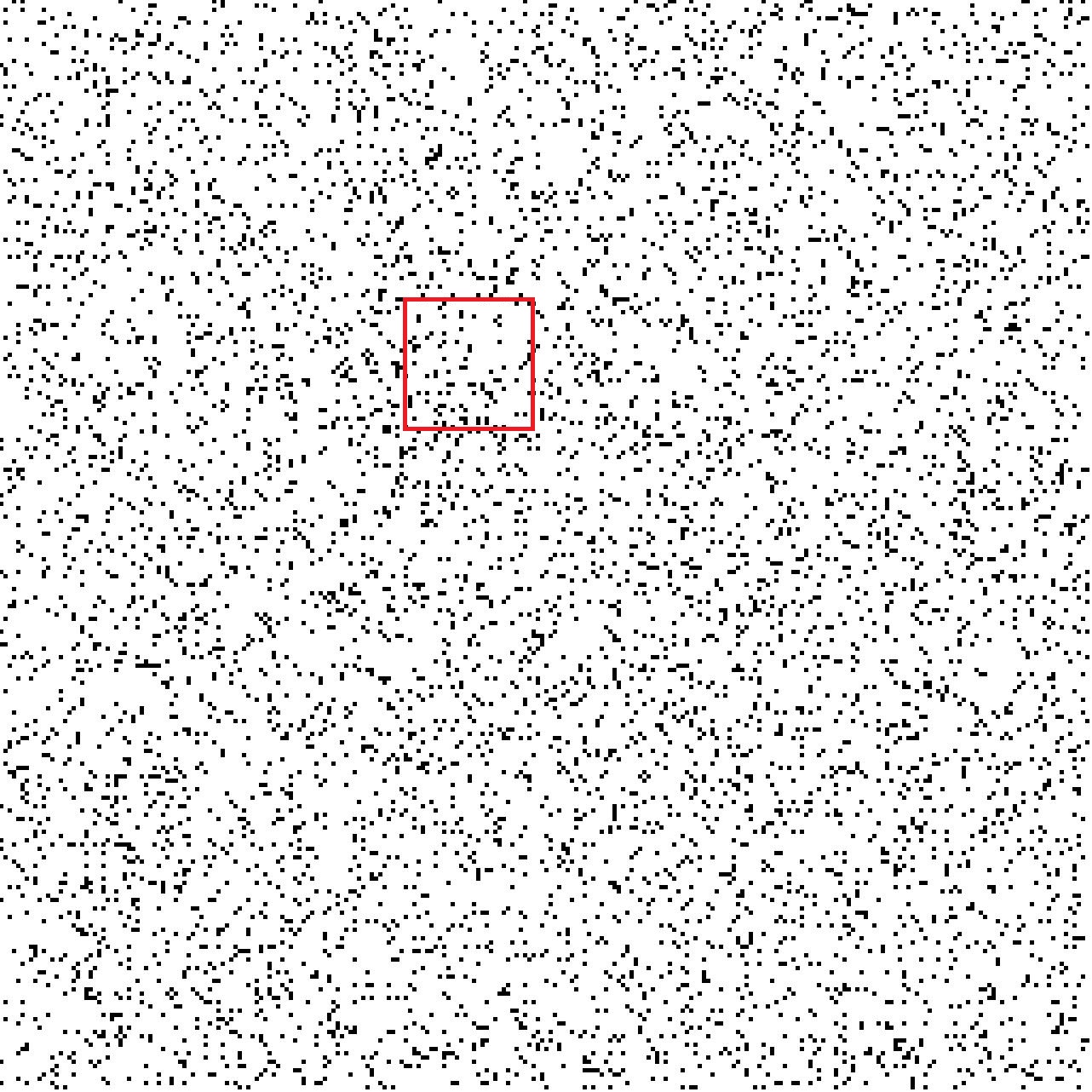}\hspace{1mm}
	\includegraphics[width=0.23\textwidth]{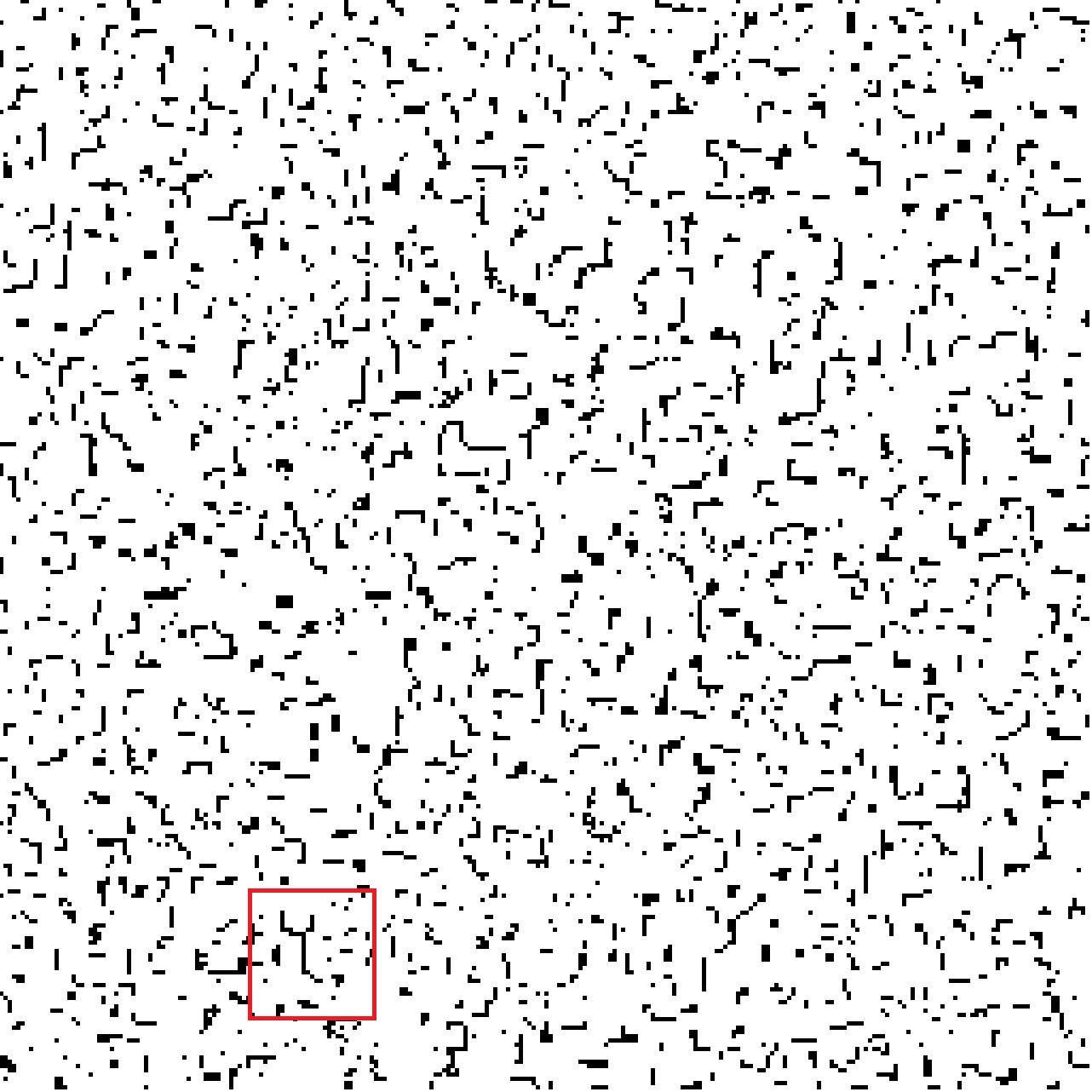}\hspace{1mm}
	\includegraphics[width=0.23\textwidth]{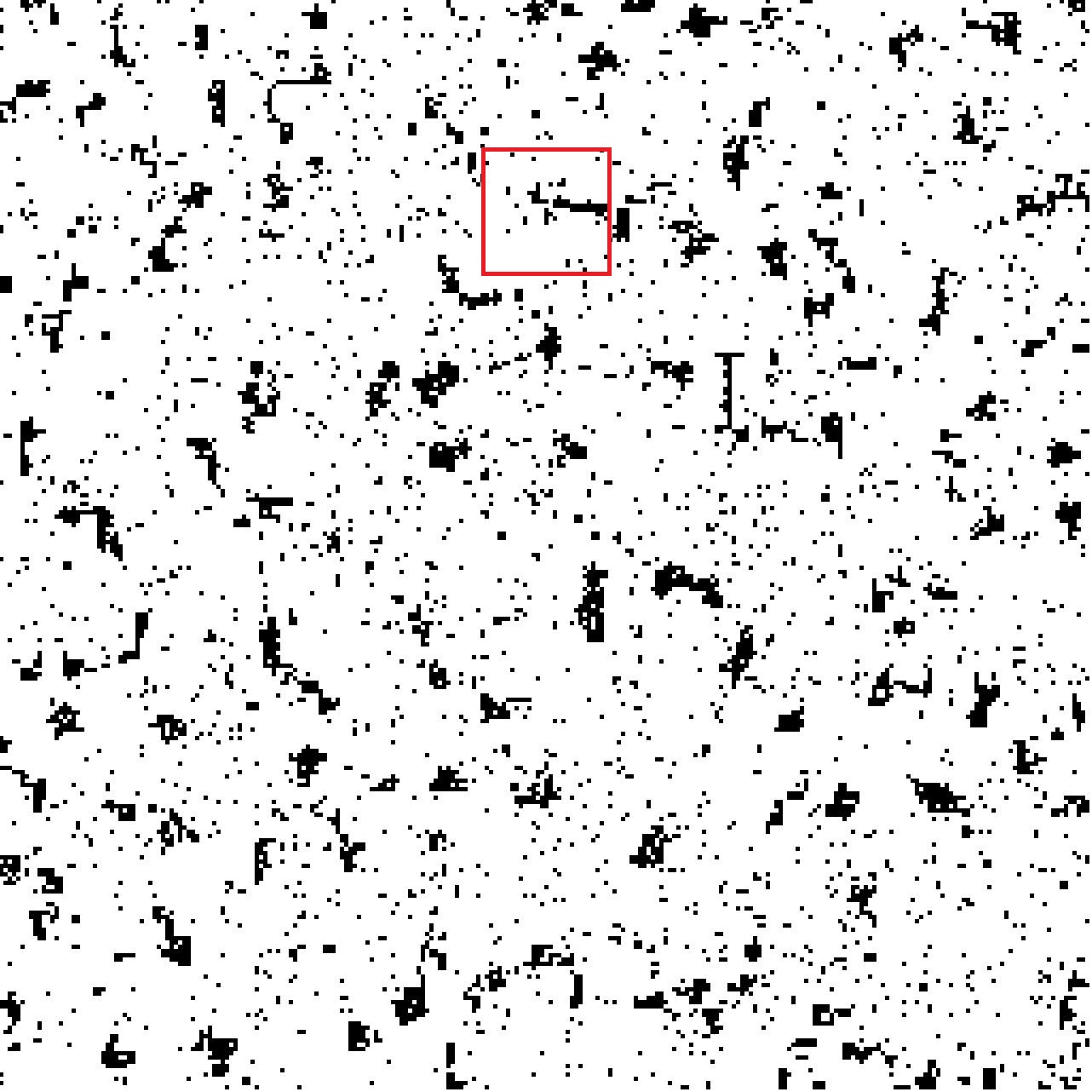}\hspace{1mm}
	\includegraphics[width=0.23\textwidth]{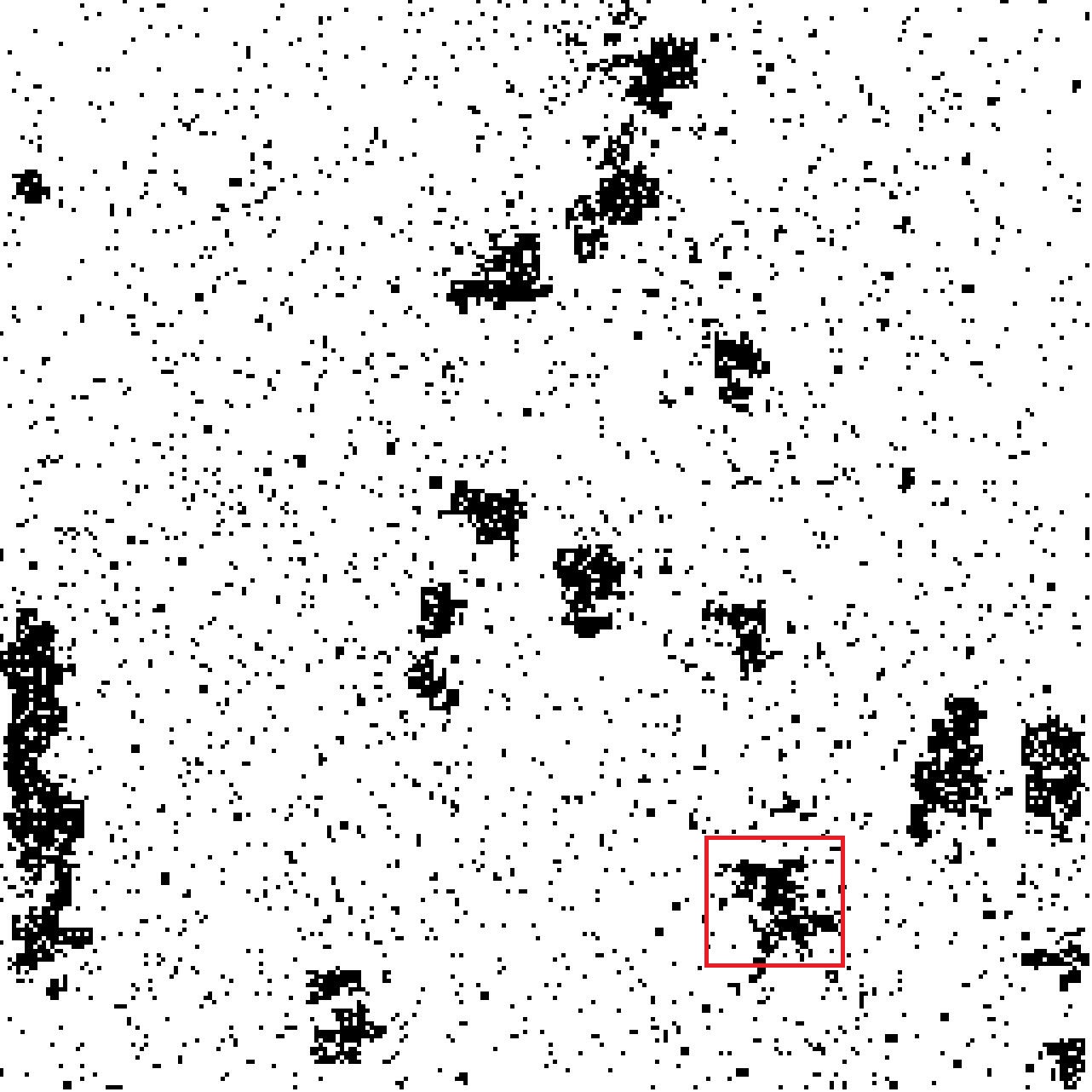}\\[2mm]
	\includegraphics[width=0.23\textwidth]{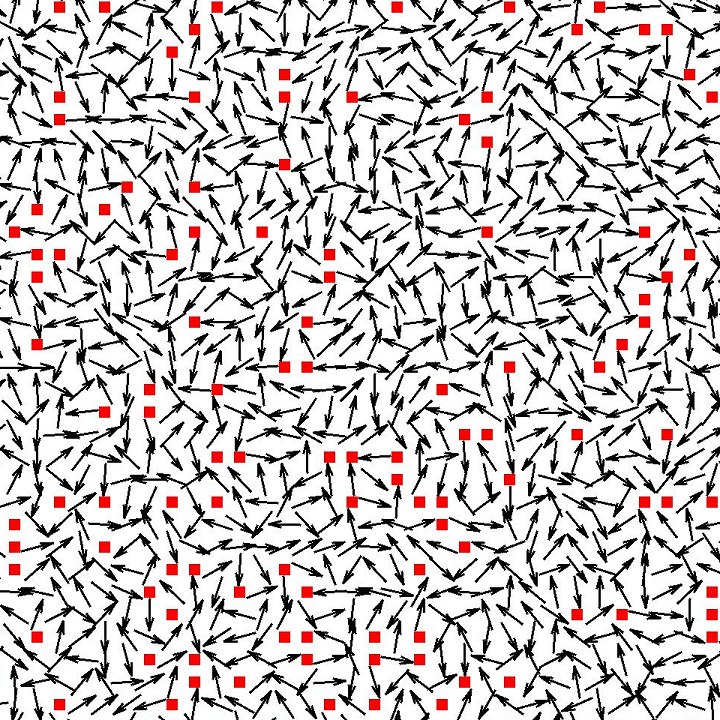}\hspace{1mm}
	\includegraphics[width=0.23\textwidth]{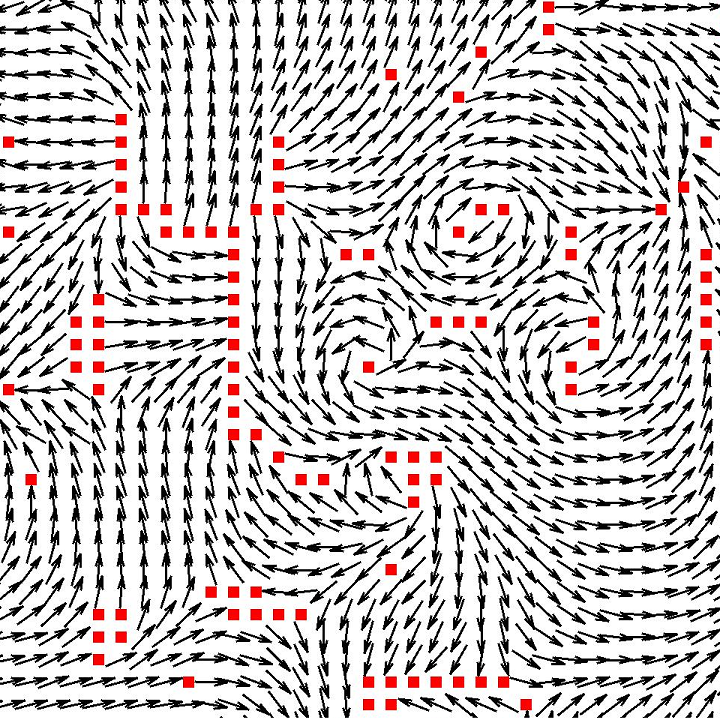}\hspace{1mm}
	\includegraphics[width=0.23\textwidth]{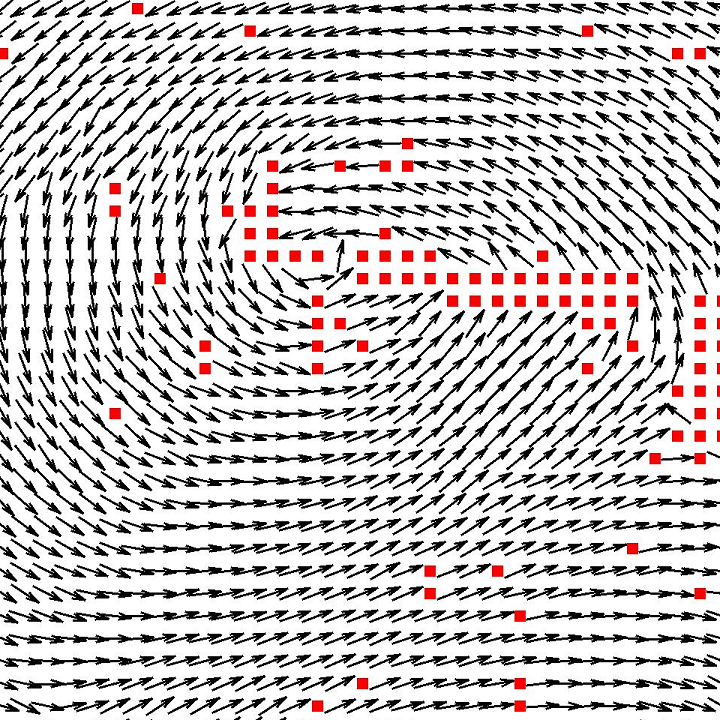}\hspace{1mm}
	\includegraphics[width=0.23\textwidth]{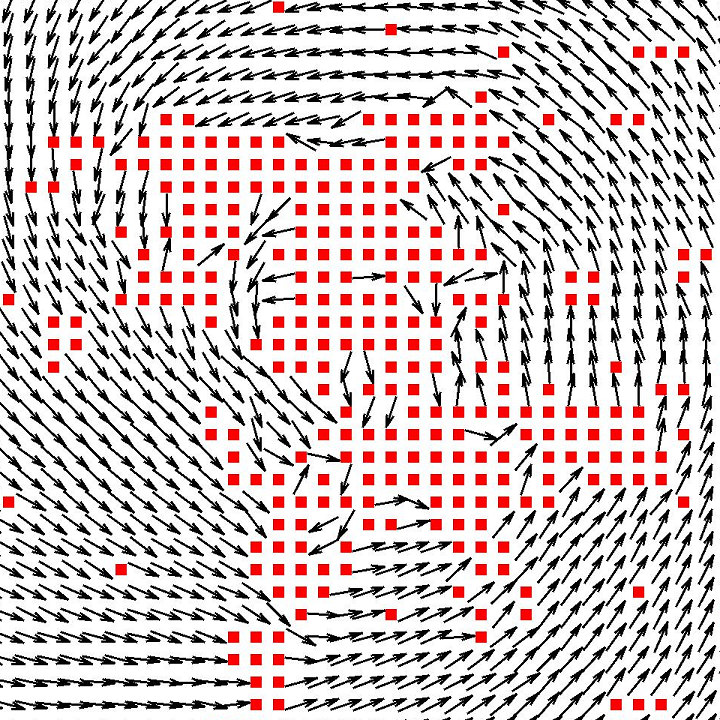}
	\caption{\label{Snapshot}
		Snapshot configuration of the system with the linear size $L = 256$
		and the spin concentration $p = 0.9$ by time $0$ (initial state), $1000$, $10000$ 
		and $50000$ $\mcs$. On top figures 
		demonstrate snapshot configuration of the 
		defects in the system. The figures below 
		show enlarged fragments system 
		configurations: arrows demonstrate spins, red squares -- defects.
	}
\end{figure*}
\begin{figure}[ht!]
	\centering
	\includegraphics[width=0.23\textwidth]{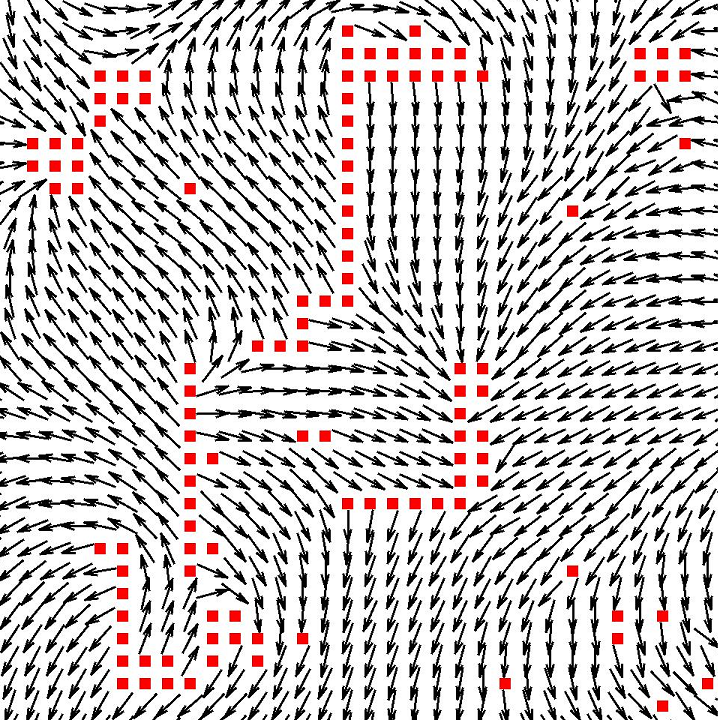}\hspace{1mm}
	\includegraphics[width=0.23\textwidth]{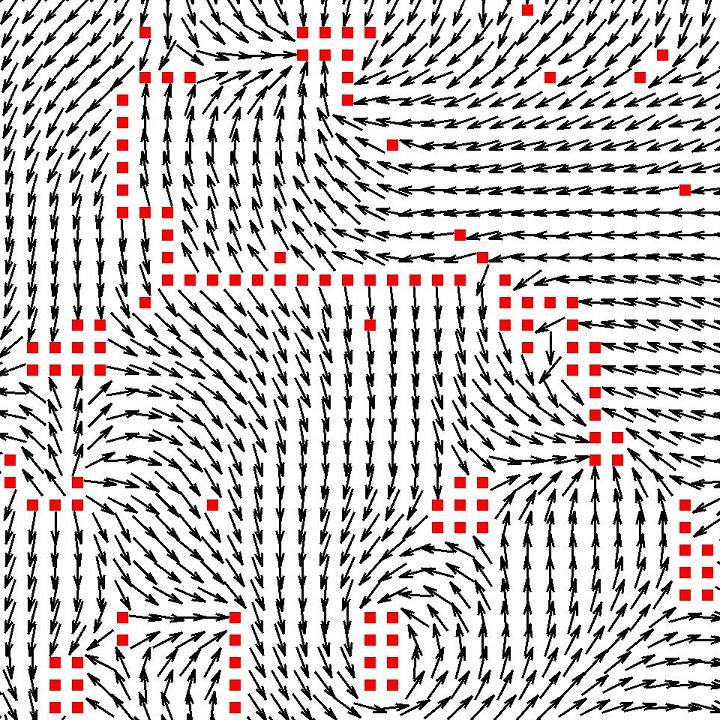}\\[2mm]
	\includegraphics[width=0.23\textwidth]{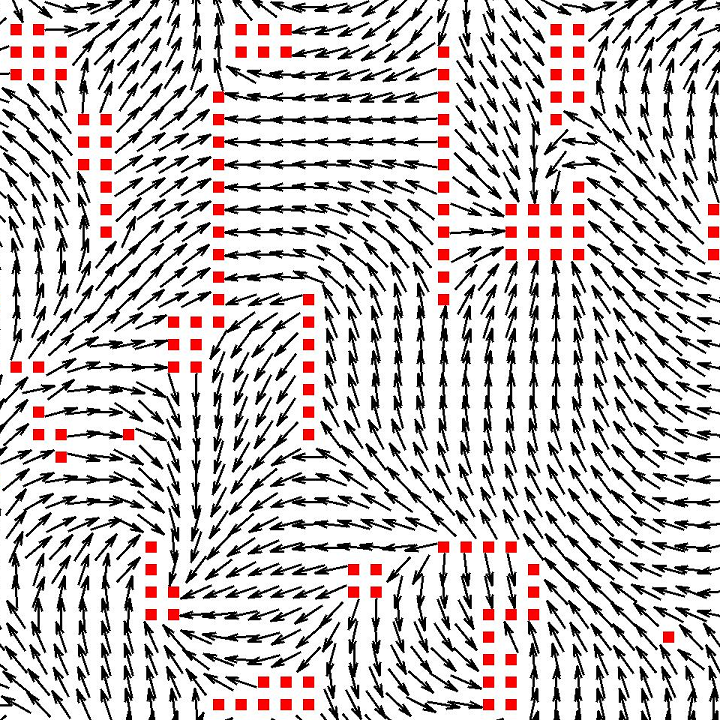}\hspace{1mm}
	\includegraphics[width=0.23\textwidth]{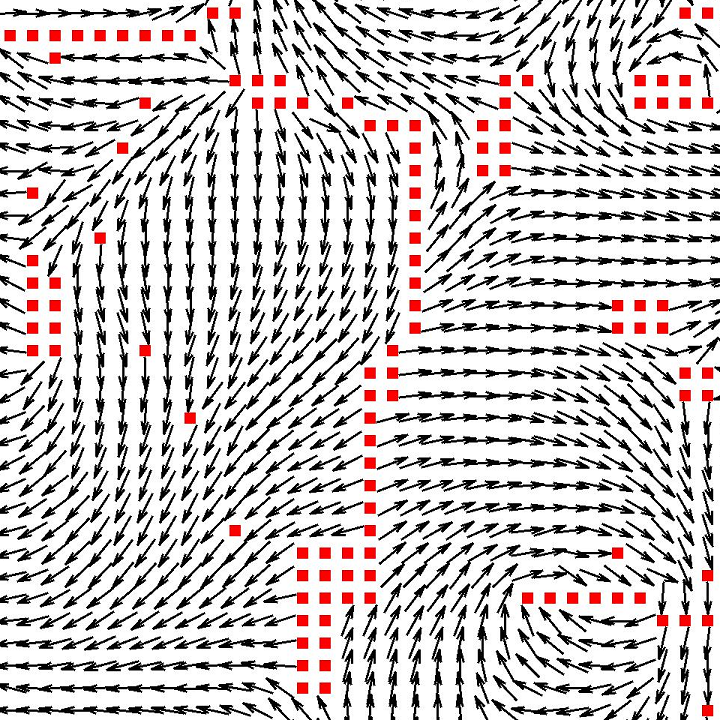}
	\caption{\label{Snapshot_qstr}
		Typical snapshot of Monte Carlo configurations which demonstrate a stripe structure of defects clusters.
		Arrows corresponds to spins, red squares corresponds to defects.
	}
\end{figure}

The XY-model is used to describe the critical 
properties of a wide range of real physical systems~\cite{Korshunov},
such as critical properties of ultra-thin magnetic films,
extensive class of ``easy plane"{} planar magnets
and critical properties of some other physical systems.
Some physical systems exhibit two-dimensional XY-like behavior under certain conditions,
such as the  frustrated Heisenberg antiferromagnets on a triangular lattice
in non-equilibrium relaxation~\cite{PPIK}.
Despite extensive research~\cite{RMP_LM,PPM}, the influence of structural disorder on 
non-equilibrium critical phenomena in the 2D XY-model is not finally resolved. 

Existing works (see refs. in%
~\cite{PPM,KapBercheHol,KapBercheHol2,Berche,PPP_2018,PPP_2015,PPPP_JOPCS_2019_CL}) are focused on the influence of quenched 
disorder on the critical behavior. 
The spin mobility and impurity annealing had to take into account for description of  granular superconductors~\cite{XY_model_ad_1,XY_model_ad_2},
high-temperature bulk superconductors~\cite{XY_model_ad_3},
two-dimensional superfluids~\cite{XY_model_ad_4}
and the superfluid transition of helium in porous media~\cite{XY_model_ad_5} by the two-dimensional XY-model.

The stripe structures can be observed in systems with mobile defects which described by  the two-dimensional XY-model \cite{ref_ad_norm_1,ref_ad_norm_2} and by other lattice models \cite{ref_ad_1,ref_ad_2,ref_ad_3,ref_ad_4,ref_ad_5} with additional special long-range interactions.

The interaction between the vortices, like a two-dimensional Coulomb gas~\cite{Berezinskii_1,Berezinskii_2,Berezinskii_book},
characterized by logarithmic dependence on distance.
However, the presence of structural disorder in ``simple" two-dimensional XY-model leads to additional interaction of vortices through a field of defects. In 2003 Pereira, \textit{et al.} \cite{Pereira} shown that for an isolated vortex there is an attracting potential with a logarithmic dependence on the distance to the defect.

Therefore, the insertion of annealed disorder, in the form of mobile defects, 
can lead to non-trivial cooperative effects in the non-equilibrium critical relaxation of 
the system, without direct inclusion of long-range potential.



In a dynamic process of pinning the vortices move to the fixed defects when 
disorder is quenched. However, the termalization of disorder leads to mutual effect -- defects 
can move to vortices cores with cluster formation. So, the vortices can aggregate defects of structure in their cores 
during non-equilibrium critical relaxation.

The initial simple Monte Carlo simulation revealed this 
effect (see FIG.~\ref{Snapshot_lc} and FIG.~\ref{Snapshot}).
It can be seen that defects are aggregated in
the vortices cores associated by vortex dynamics, but the clusters of aggregated defects disappeared in equilibrium state. The disorder is gradually thermalized with spin-wave dynamics when vortices with a opposite topological charges annihilated. 
The typical snapshot of Monte Carlo configurations 
clear demonstrates that clusters of aggregated defects has
a stripe structure at short times $t$ (see FIG.~\ref{Snapshot} and, especially, FIG.~\ref{Snapshot_qstr}).

It should be noted that the revieled stripes structures are significantly different from the stripes 
identified in \cite{ref_ad_norm_1,ref_ad_norm_2,ref_ad_2,ref_ad_3}).
In the present case structures has essentially non-equilibrium nature and 
their existence is supported by non-equilibrium vortex dynamics.
On large dynamic time scales our stripe clusters collapse with forming a clumps structures~\cite{ref_ad_norm_1,ref_ad_norm_2}
(see FIG.~\ref{Snapshot} for cases $10000$ and $50000$ $\mathrm{MCS/s}$). All these effects disappeared in the equilibrium state.



It was shown~\cite{Bray} that the canonical dynamical scaling is violated in non-equilibrium 
vortex dynamics in pure 2D XY-model and correlation length behavior becomes $\xi \sim (t / \ln t)^{1/z}$
instead of $\xi \sim t^{1/z}$, with dynamical critical exponent $z = 2$. The change is due to the friction of the 
vortices in the process of motion with the friction constant $\gamma(R) \sim \ln(\xi/a)$~\cite{Bray}.
Similar behavior occurs in other models 
(frustrated Heisenberg antiferromagnets on a triangular lattice~\cite{PPIK}) 
with similar non-equilibrium vortex dynamics.
The presence of defects aggregation can change this relation and introduce new dynamical scaling behavior.
The presence of defects aggregation can change this relation and introduce new dynamical scaling behavior.


The Hamiltonian of the system in this work was chosen in the form
\begin{equation}
	H = -\frac{J}{2} \sum_{i, j} p_{i} p_{j} \mathbf{S}_{i} \mathbf{S}_{j},
\end{equation}
where $\mathbf{S}_{i}$ is a classical planar spin which is associated with $i$-node of
square lattice with the linear size $L$, $J>0$ is exchange integral, 
$p_{i}$ is occupation number of $i$-node,  
$\sum_{i, j}$ provide a summation over all pairs of the nearest neighbors.
Defects on the lattice are distributed uniformly at time $t = 0$ with probability 
$P(p_{i}) = (1 - p) \delta(p_{i}) + p \delta(1 - p_{i})$, where $p$ is a spin concentration,
i.e. $c = 1 - p$ is a concentration of defects. 
The temperature $T$ of the system is measured in units of the exchange integral $J/k_b$.

The simulation of critical spin dynamics of the two-dimensional XY-model was realized with the Metropolis algorithm.
The sequence of states defined by the Metropolis algorithm according 
to the transition probability between neighboring configurations, forms a Markov chain.
The evolution of the non-equilibrium distribution function $P_n(t)$ 
can be presented in the form of the master kinetic equation with the
transition probability $W(n \rightarrow m) = \min \left[1.0, \exp(-\Delta E_{nm}/T)\right]$.
The dynamics implemented by the Metropolis algorithm corresponds to the 
dynamical model A~\cite{Hohenberg_1977,Folk}.
It was shown~\cite{PP_FMM} that the Metropolis algorithm correctly describes 
a non-equilibrium critical properties of the two-dimensional XY-model.
To include the mobility of defects, the elementary step of the Metropolis algorithm was modified. 
If a randomly selected node is occupied by a spin, then classical elementary step is produced. 
But if node is occupied by a defect, then an attempt is made to swap this defect with random 
neighboring spin. Similarly, the calculation of $\Delta E_{nm}$ is carried out and 
elementary step of Metropolis algorithm is used.
As a time unit, we used the Monte Carlo step per spin ($\mcs$) 
corresponding to $L^{2}$ spin flips or defects moves per unit time.
The study was carried out for the spin concentrations $p=0.9$, $0.8$, $0.7$ and $0.6$.

\begin{figure*}[ht!]
	\centering
	\includegraphics[width=0.41\textwidth]{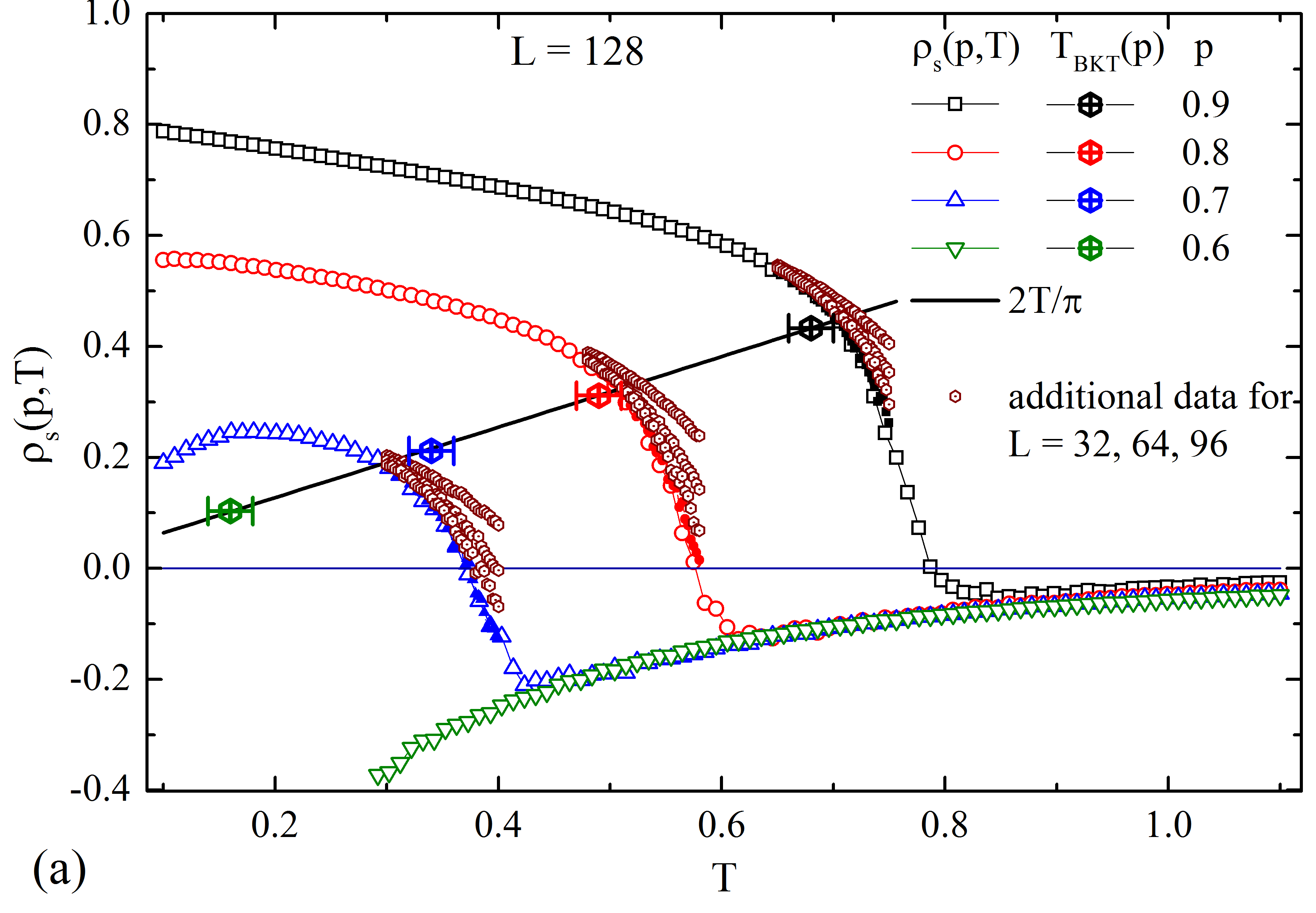}\hspace{12mm}
	\includegraphics[width=0.41\textwidth]{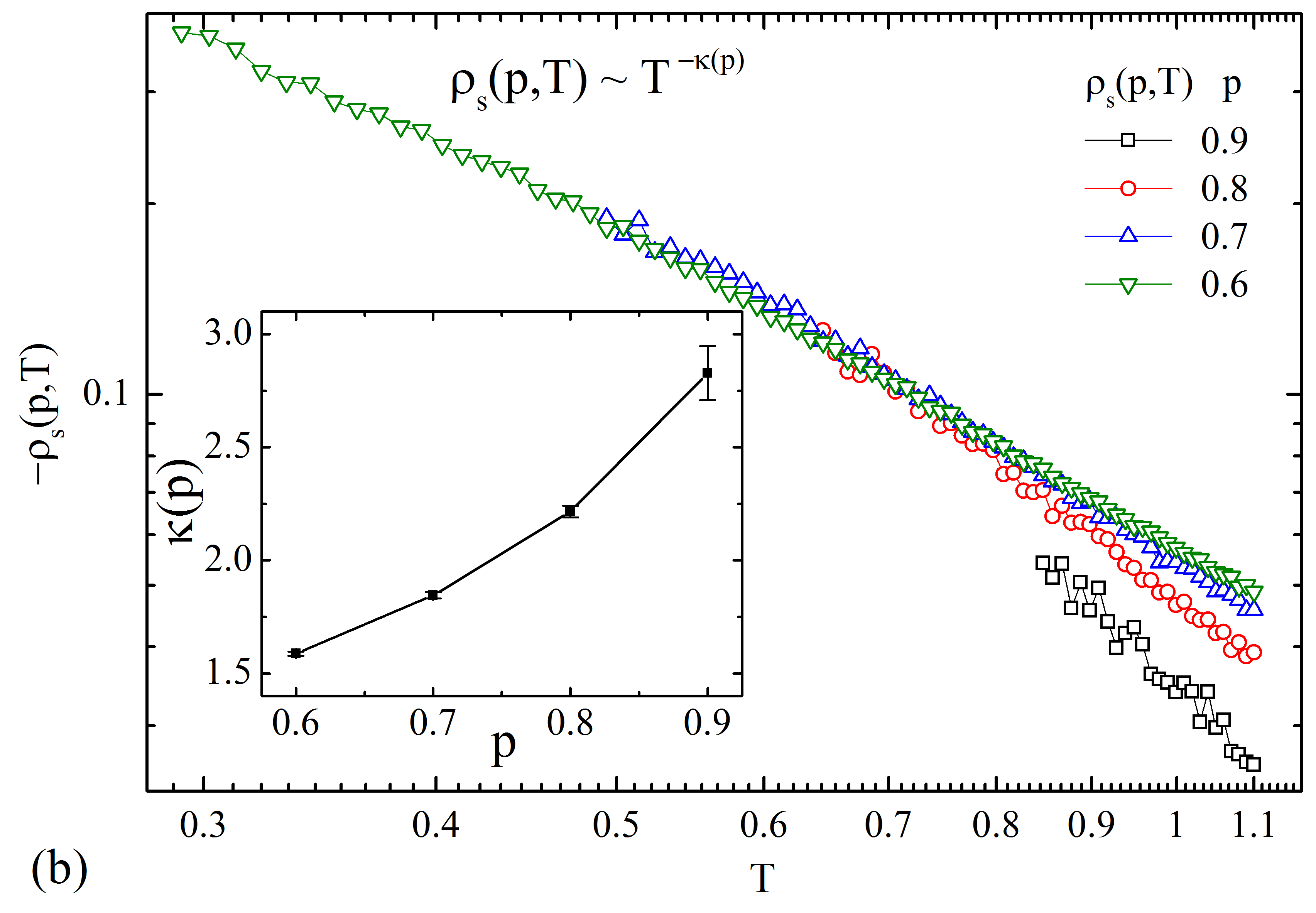}
	\caption{\label{rho_s}
		The temperature dependence of transverse stiffness $\rho_{\rm s}(p,T)$
		for the system with spin concentrations $p$ $=$ $0.9$, $0.8$, $0.7$, $0.6$ and
		linear size $L = 128$ (also in the figure are shown small areas with
		$L$ $=$ $32$, $64$, $96$ to demonstrate the effects of the finite linear size). 
		In fig. (a) in addition are shown the values $\TBKTp$
		and linear dependence $2T/\pi$. In fig. (b) is shown temperature dependence 
		of negative value of transverse stiffness $\rho_{\rm s}(p,T)$, and
		it demonstrates power dependence $\rho_{\rm s}(p,T) \sim T^{-\kappa(p)}$.
		The inset shows concentration dependence of power exponent $\kappa(p)$.
	}
\end{figure*}

The investigation of the equilibrium critical properties was carried out for the system
with linear sizes $L=32$, $64$, $96$ and $128$. To reduce the relaxation time, 
the following approach was used: the final state of the simulated system at temperature $T$
was used as the initial state of system at temperature $T + \delta T$ with $\delta T \ll T$.
For thermalization it was used $10^{6}$ $\mcs$ for initial temperature and $10^{4}$ $\mcs$ for 
subsequent temperatures. The averaging was carried out over the $10^{5}$ $\mcs$ and
$1000$ different initial impurity configurations.

The investigation of non-equilibrium critical relaxation was carried out for system
with linear sizes $L$ $=$ $32$, $64$, $128$ and $256$ and observation time $10^{4}$ $\mcs$ 
for dynamic dependencies and with $10$ different linear sizes $L$ $=$ $24$, 
$32$, $\dots$, $96$ and observation time $10^{3}$ $\mcs$ for dynamical scaling investigation.
We used $10000$ and $150000$ different initial impurity configurations
for dynamic dependencies and for dynamical scaling investigation correspondingly.

An important feature of non-equilibrium annealing of structure disorder is the clustering of defects.
A structure of defects clusters is determined by the Hoshen-Kopelman 
algorithm~\cite{HKAlg_1,HKAlg_2,HKAlg_3}.
For each time step $t$, it were searched all defects clusters on the lattice and
it were calculated a sizes $S_{k}$ for each $k$ cluster.
The size $S_{k}$ corresponds to amount of defects in $k$ cluster.
The calculations were made for the sizes of the largest clusters $S_{\rm m}(p,T;t)$
and the averaged size of clusters of defects $S_{\rm av}(p,T;t)$
\begin{equation} \hspace*{-2mm}
\begin{array}{rcl}
S_{\rm m}(p,T;t) &=& \left[\left\langle \max\limits_{k} S_{k}(p,T;t) \right\rangle\right], \\[3mm]
S_{\rm av}(p,T;t) &=& \left[\left\langle \frac{1}{K} \sum\limits_{k} S_{k}(p,T;t) \right\rangle\right],
\end{array}\hspace*{-1mm} k=1\ldots K,
\end{equation}
where $K = K(p,T;t)$ is the full amount of clusters on the lattice,
the brackets $\langle\ldots\rangle$ and $[\ldots]$ corresponds to 
the statistical averaging over spin and impurity configurations, consequently. 
In the process of non-equilibrium critical relaxation, $S_{\rm m}(p,T;t)$ and $S_{\rm av}(p,T;t)$ characterized by time dependence.
Equilibrium values of $S_{\rm m}(p,T)$ and $S_{\rm av}(p,T)$ can be obtained by additional 
averaging over independent time sampings.


We provide massive Monte Carlo calculation for precise determination of the temperatures of BKT phase transition $\TBKTp$.
We used Binder's cumulants
and ratio of a correlation functions \cite{Berche,PPP_JOPCS_2014}.
So, the critical temperatures for different spin concentrations $p$ take on the following values: 
$\TBKT(p=0.9) = 0.68(1)$, $\TBKT(p=0.8) = 0.49(2)$, $\TBKT(p=0.7) = 0.34(2)$ and $\TBKT(p=0.6) = 0.16(2)$.
It is well known that annealed structural disorder 
does not significantly affect the equilibrium critical behavior~\cite{MaS}, 
and the calculated temperatures $\TBKTp$ for 
systems with mobile disorder agree with values of temperatures for quenched disordered system~\cite{Berche} for all considered spin concentrations $p$ $=$ $0.9$, $0.8$, $0.7$ and $0.6$.

\begin{figure*}[ht!]
	\centering
	\includegraphics[width=0.41\textwidth]{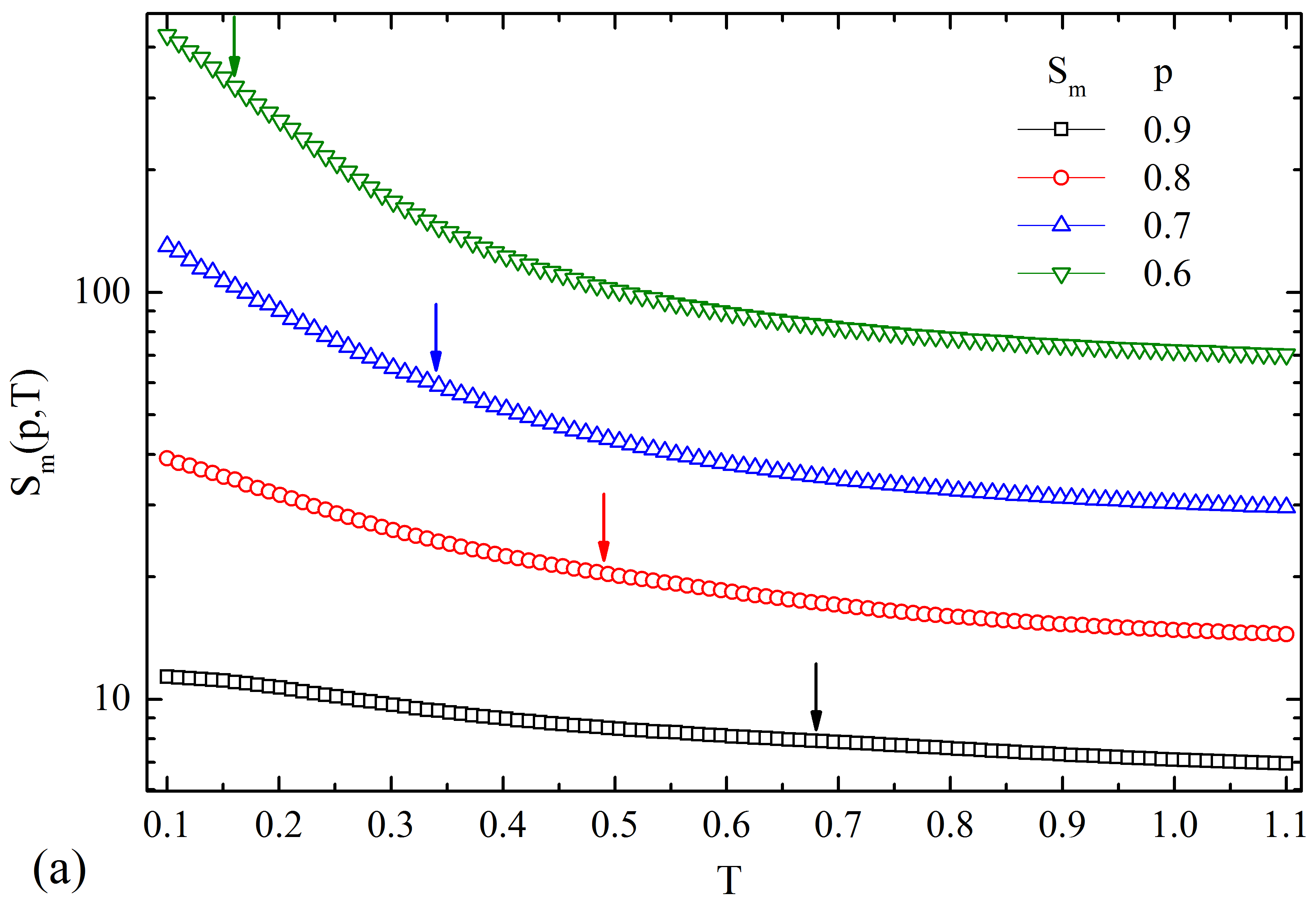}\hspace{12mm}
	\includegraphics[width=0.41\textwidth]{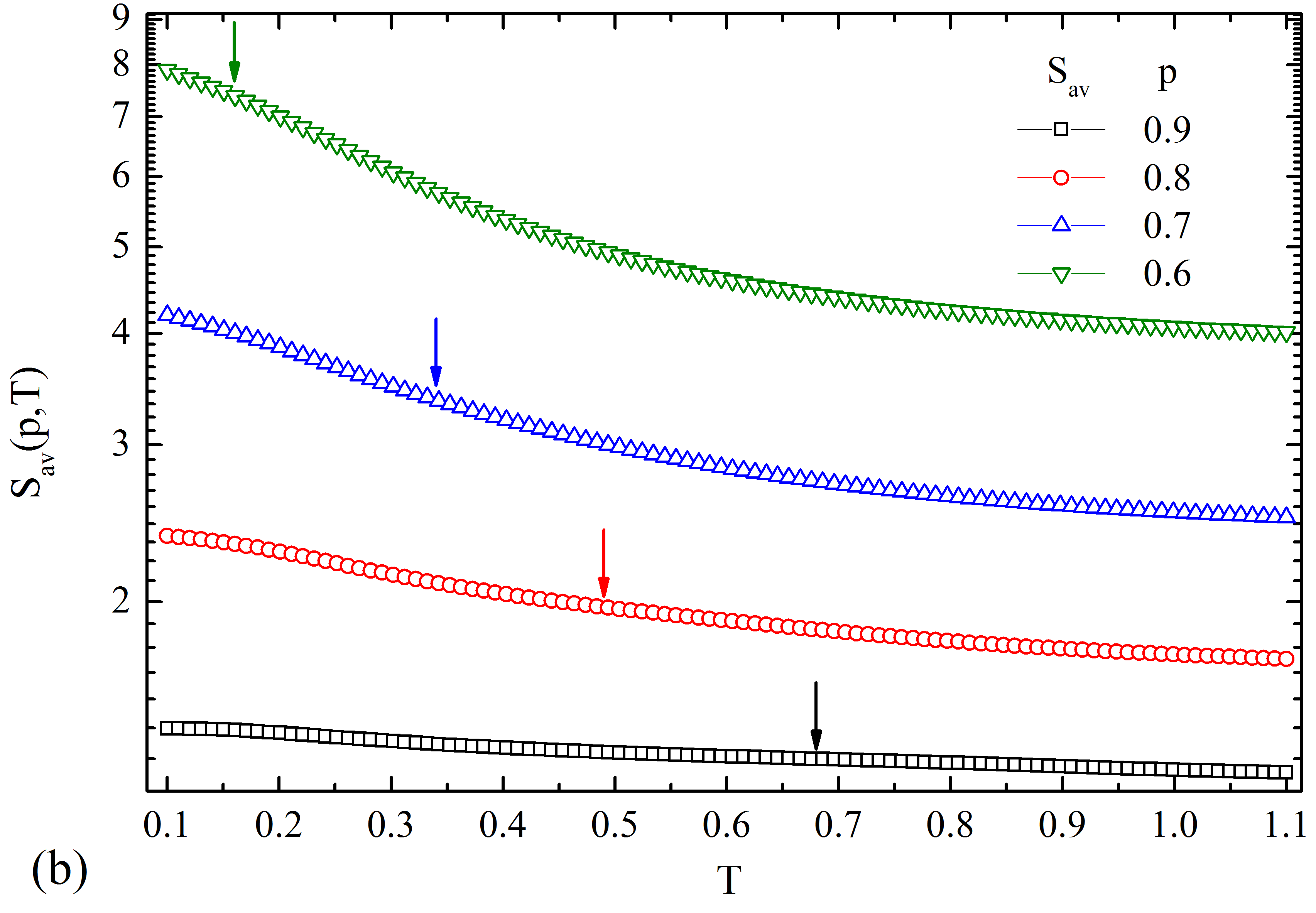}
	\caption{\label{CS_eq}
		The temperature dependence of size of the largest cluster of defects $S_{\rm m}(p,T)$ (a)
		and average size of clusters of defects $S_{\rm av}(p,T)$ (b)
		for system with spin concentrations $p$ $=$ $0.9$, $0.8$, $0.7$, $0.6$ and
		linear size $L = 128$. Vertical arrows indicate the position of $\TBKTp$.}
\end{figure*}

It should be noted that the spin percolation threshold 
$p_{\rm c} \simeq 0.592745(2)$~\cite{Percolation}
and $p = 0.6$ is sufficiently close to it. 
The dependencies of $S_{\rm m}(p,T)$ and $S_{\rm av}(p,T)$ which are presented on FIG.~\ref{CS_eq} let us possibility to conclude that the defects are distributed on lattice nonuniformly in equilibrium state.
In case of nonuniformly distributed defects a problem of the phase transition break near the spin percolation threshold becomes not trivial and it is not discussed here in detail.


The most important property of the low-temperature phase $T < \TBKT$ 
is the non-zero value of the transverse stiffness $\rho_{\rm s}$%
~\cite{Berezinskii_1,Berezinskii_2,Berezinskii_book,Kosterlitz_Thouless,Kosterlitz,Korshunov}.
In this case, at the point $T = \TBKT$, 
the magnitude of the transverse stiffness $\rho_{\rm s}$ is equal to zero.
We perform calculation of transverse stiffness~\cite{Stiff}:
\begin{equation}\hspace*{-2mm}
\begin{array}{rl}
\rho_{\rm s} = \frac{1}{2 pL^{2}}&\hspace*{-3mm}\Big[\Big<\sum\limits_{i, j} p_{i} p_{j} \cos(\varphi_{i} - \varphi_{j}) \\
			  &\hspace*{-3mm} - \frac{1}{T}\sum\limits_{l} \sum\limits_{i, j} p_{i} p_{j} \sin(\varphi_{i} - \varphi_{j}) 
				\mathbf{e}_{i j} \mathbf{x}_{l} \Big>\Big],
\end{array}
\end{equation}
where $\varphi_{i}$ is the phase of a spin $\mathbf{S}_{i}$,
$\mathbf{e}_{i j}$ is the unit vector from $i$-node to $j$-node
and $\mathbf{x}_{l} : \big[ \mathbf{x}_1=(1, 0);\  \mathbf{x}_2=(0, 1)\big]$.
The first term corresponds to the value of the energy 
and coincides with the Hamiltonian of the system.
In this work, we calculate the temperature dependencies of transverse stiffness $\rho_{\rm s}(p,T)$
for the system with spin concentrations $p$ $=$ $0.9$, $0.8$, $0.7$, $0.6$ and
linear size $L = 128$ (see FIG.~\ref{rho_s}).
It was revealed temperature intervals where the transverse stiffness has negative values and 
temperature behavior of $\rho_{\rm s}(p,T)$ can be described by power law 
\begin{equation}
	\rho_{\rm s}(p,T) \sim T^{-\kappa(p)},
\end{equation}
where exponent $\kappa(p)$ depends on $p$ only.
It is easy to see from data which presented on FIG~\ref{rho_s}, that 
the $\rho_{\rm s}(p,T)$ changes abruptly near the $\TBKTp$.
The additional data for $L$ $=$ $32$, $64$, $96$, which are presented on FIG~\ref{rho_s}, 
demonstrates that values of changing $\rho_{\rm s}(p,T)$ near $\TBKTp$ are increased for large $L$.

\begin{figure*}[ht!]
	\centering
	\includegraphics[width=0.41\textwidth]{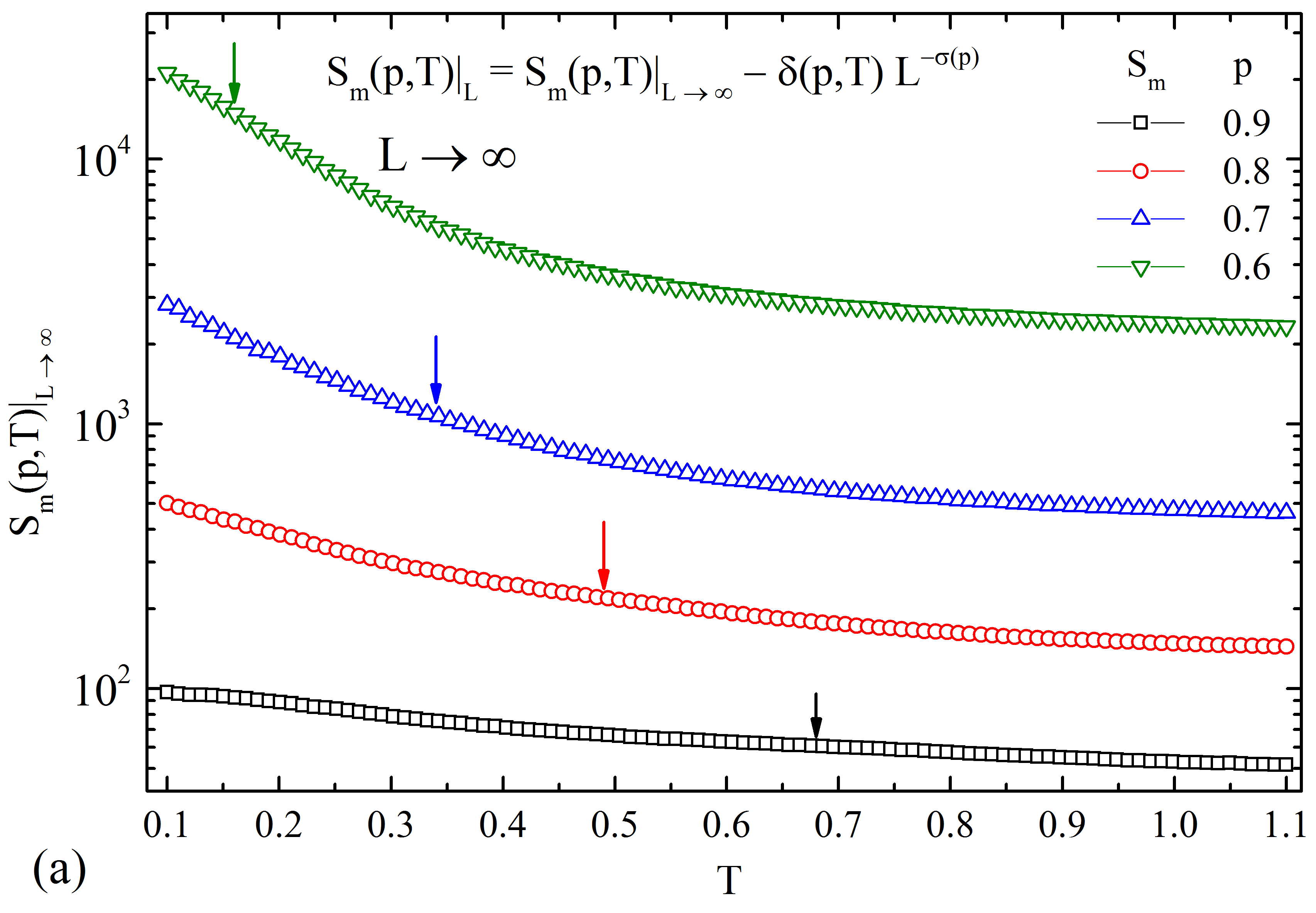}\hspace{12mm}
	\includegraphics[width=0.41\textwidth]{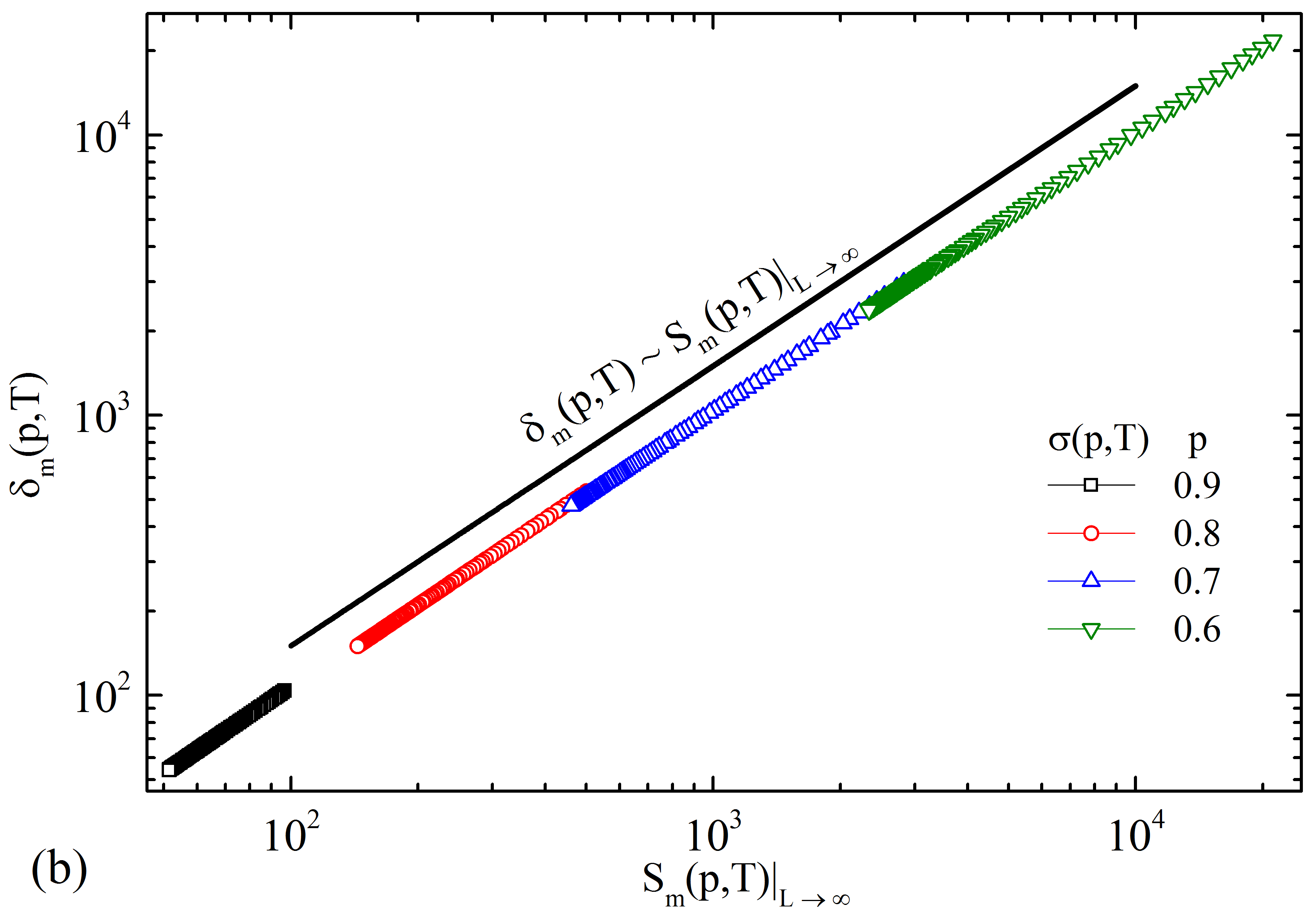}
	\caption{\label{CS_eq_scale}
		The temperature dependence of extrapolation to $L \rightarrow \infty$ 
		of size of the largest cluster of defects $S_{\rm m}(p,T) |_{L \rightarrow \infty}$ (a) and
		parametric dependence $\delta(p,T)$ vs $S_{\rm m}(p,T) |_{L \rightarrow \infty}$ (b)
		for system with spin concentrations $p$ $=$ $0.9$, $0.8$, $0.7$, $0.6$. 
		Vertical arrows indicate the position $\TBKTp$.}
\end{figure*}

\begin{figure*}[ht!]
	\centering
	\includegraphics[width=0.41\textwidth]{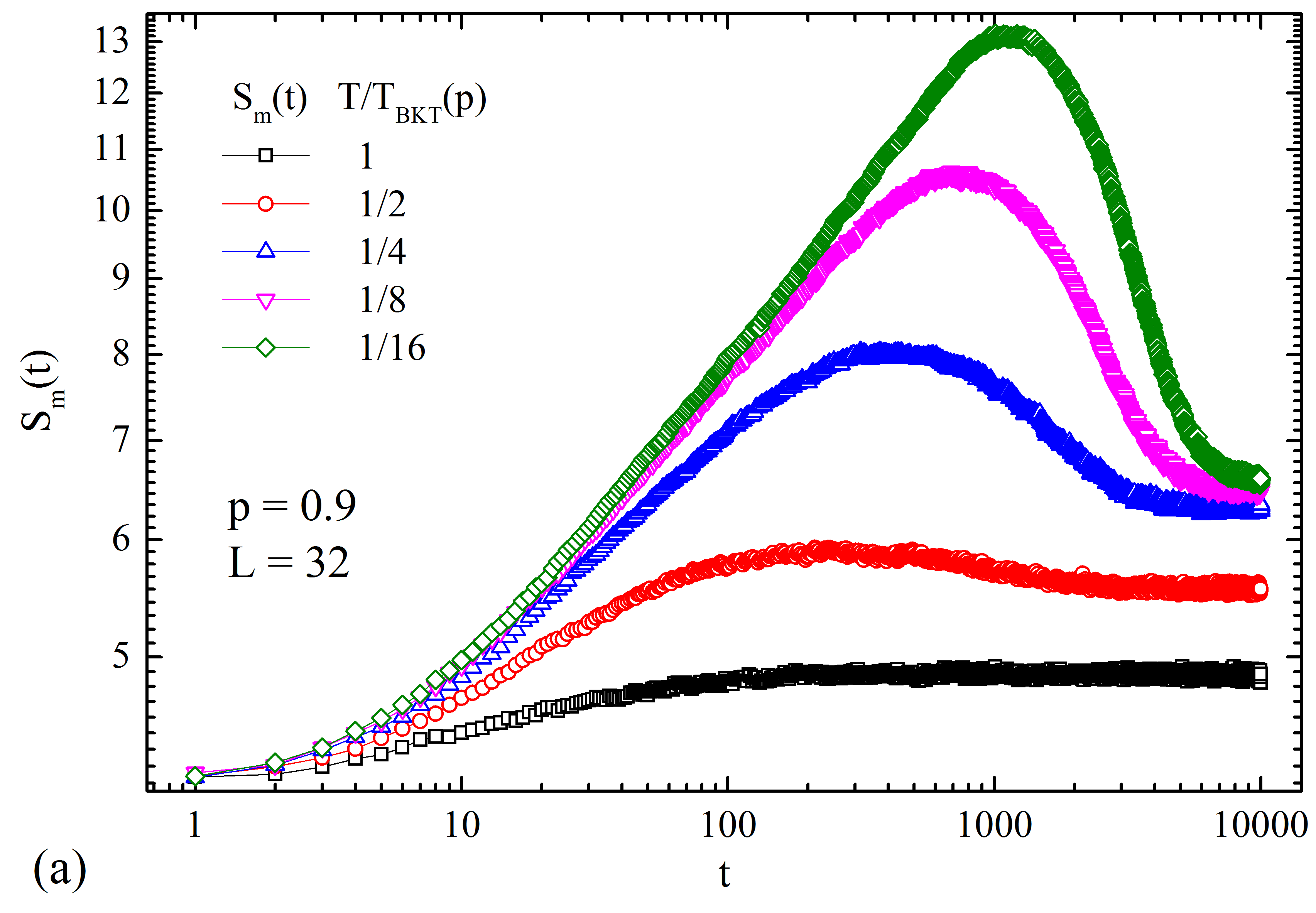}\hspace{12mm}
	\includegraphics[width=0.41\textwidth]{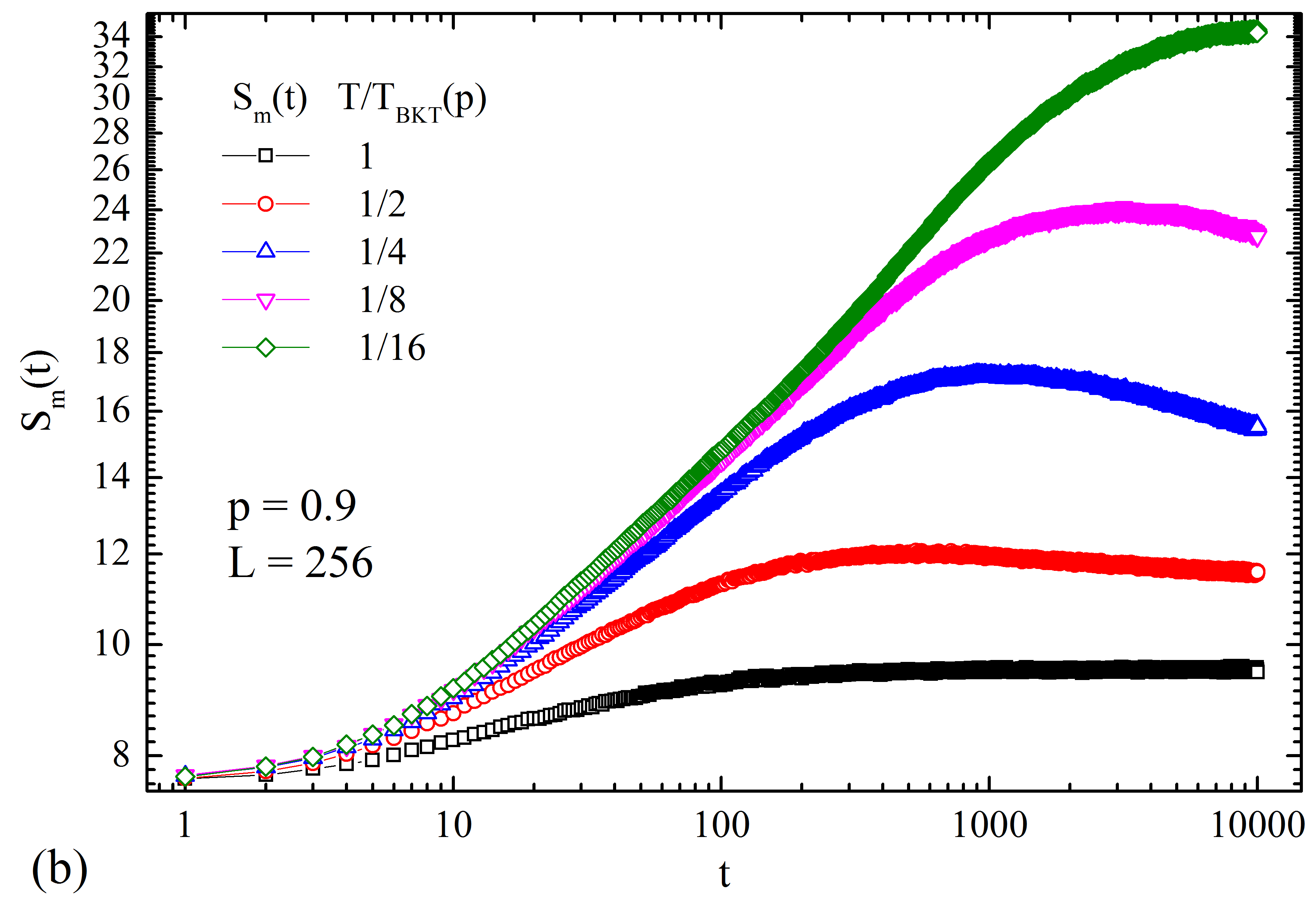}\\
	\includegraphics[width=0.41\textwidth]{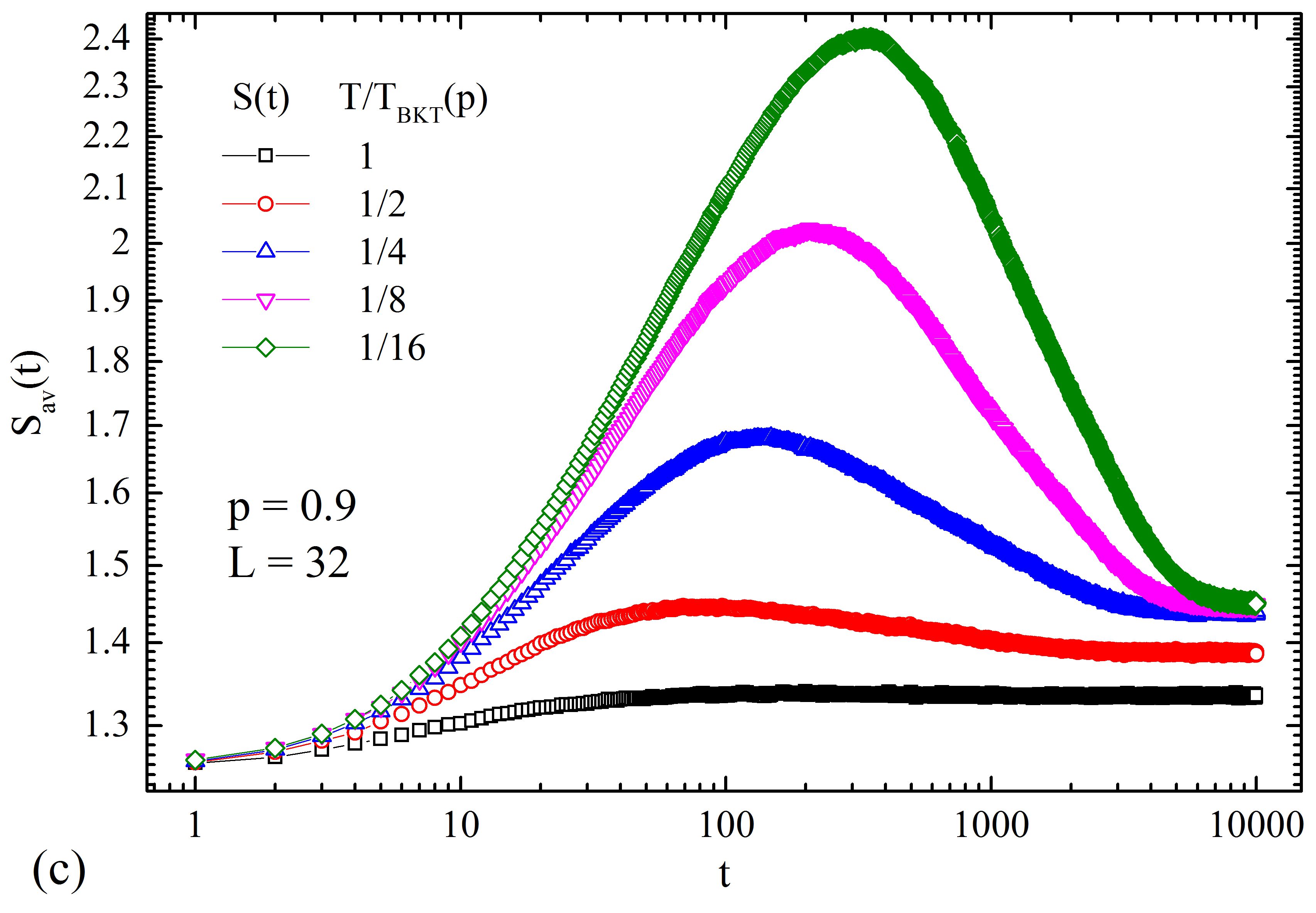}\hspace{12mm}
	\includegraphics[width=0.41\textwidth]{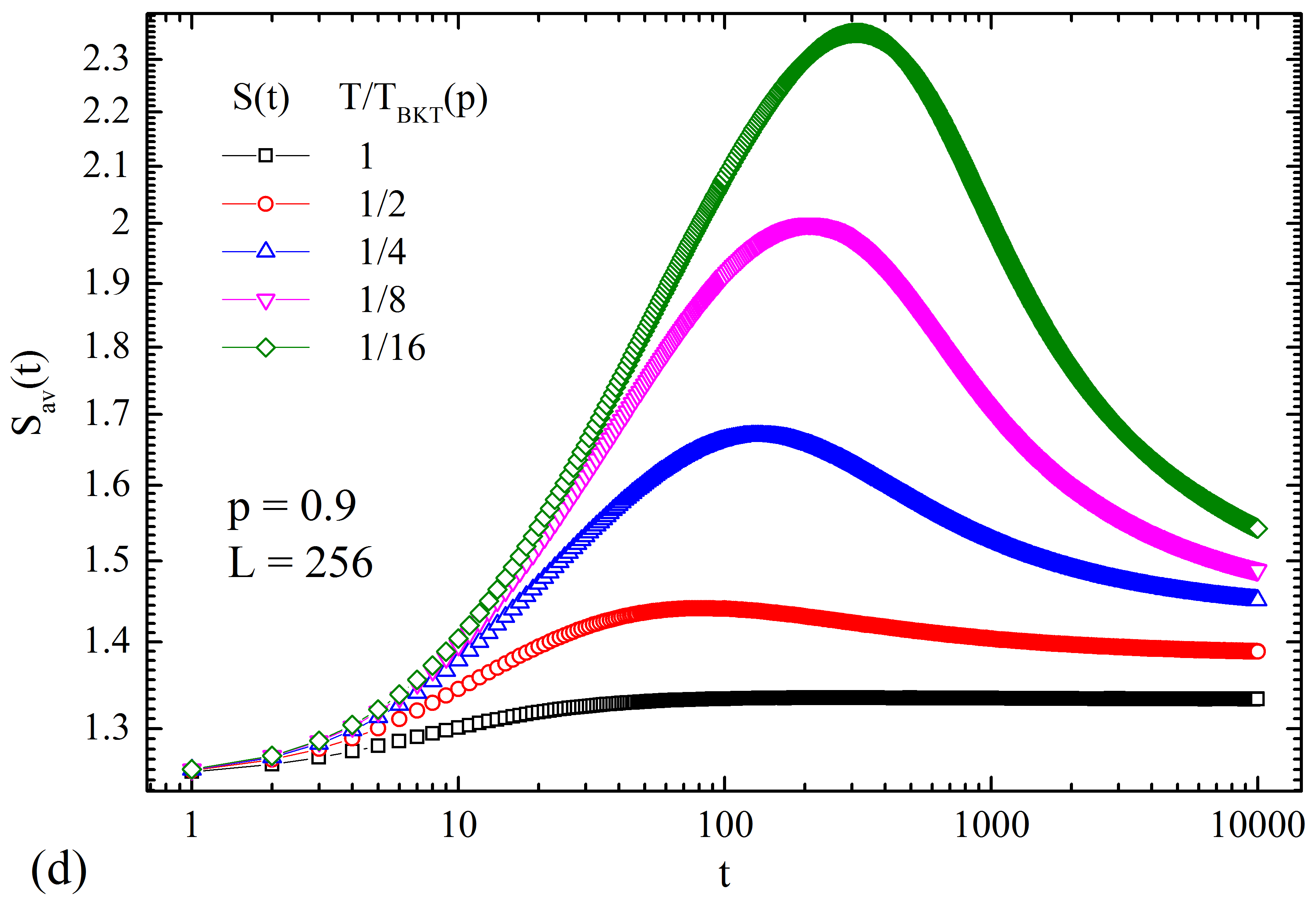}
	\caption{\label{CS_noneq_p09}
		The dynamic dependencies of size of the largest cluster of defects $S_{\rm m}(t)$ (a, b)
		and averaged size of clusters of defects $S_{\rm av}(t)$ (c, d) for
		system with spin concentration $p = 0.9$ and linear sizes $L = 32$ (a, c)
		and $L = 256$ (b, d).
	}
\end{figure*}
\begin{figure*}[ht!]
	\centering
	\includegraphics[width=0.41\textwidth]{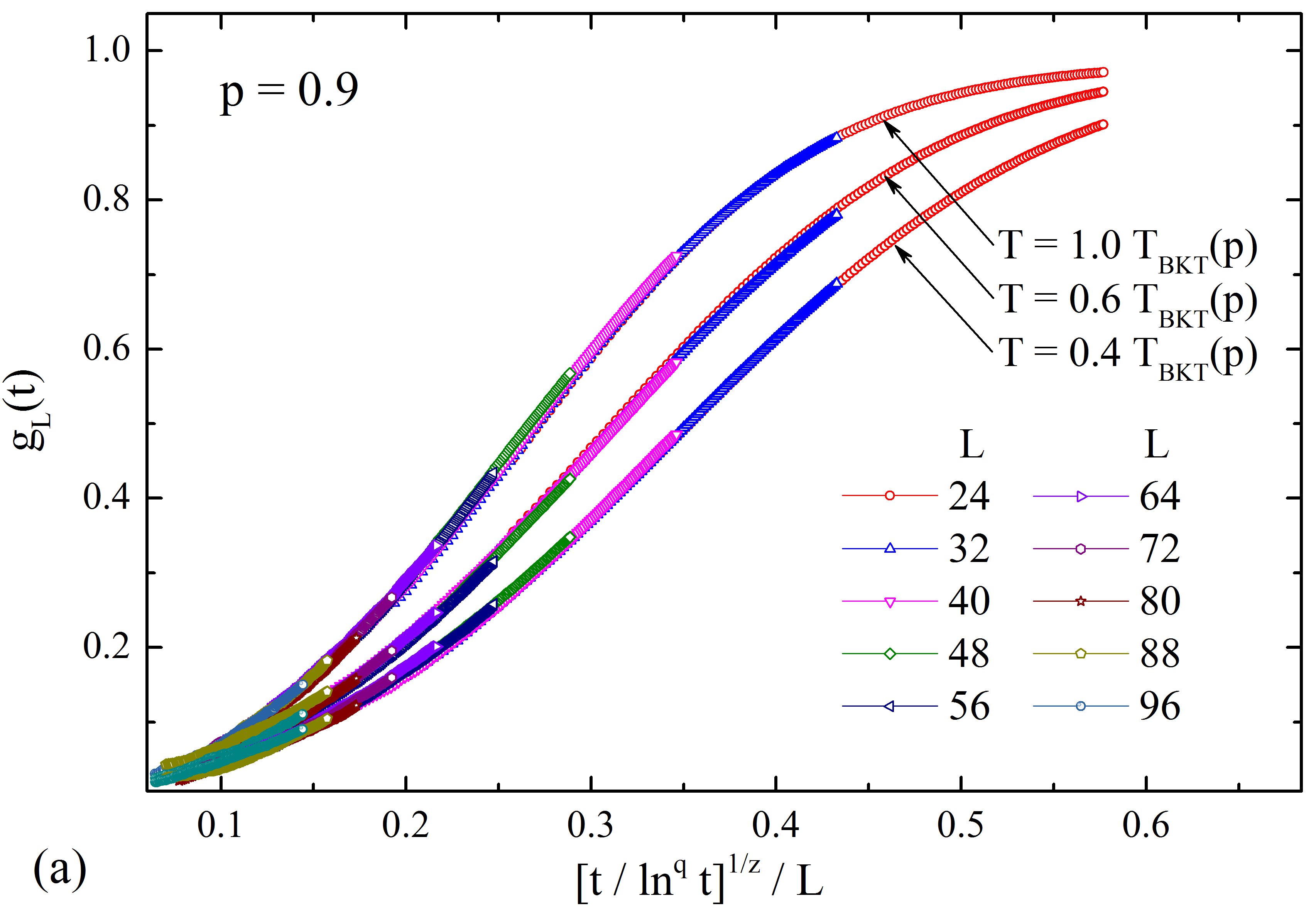}\hspace{12mm}
	\includegraphics[width=0.41\textwidth]{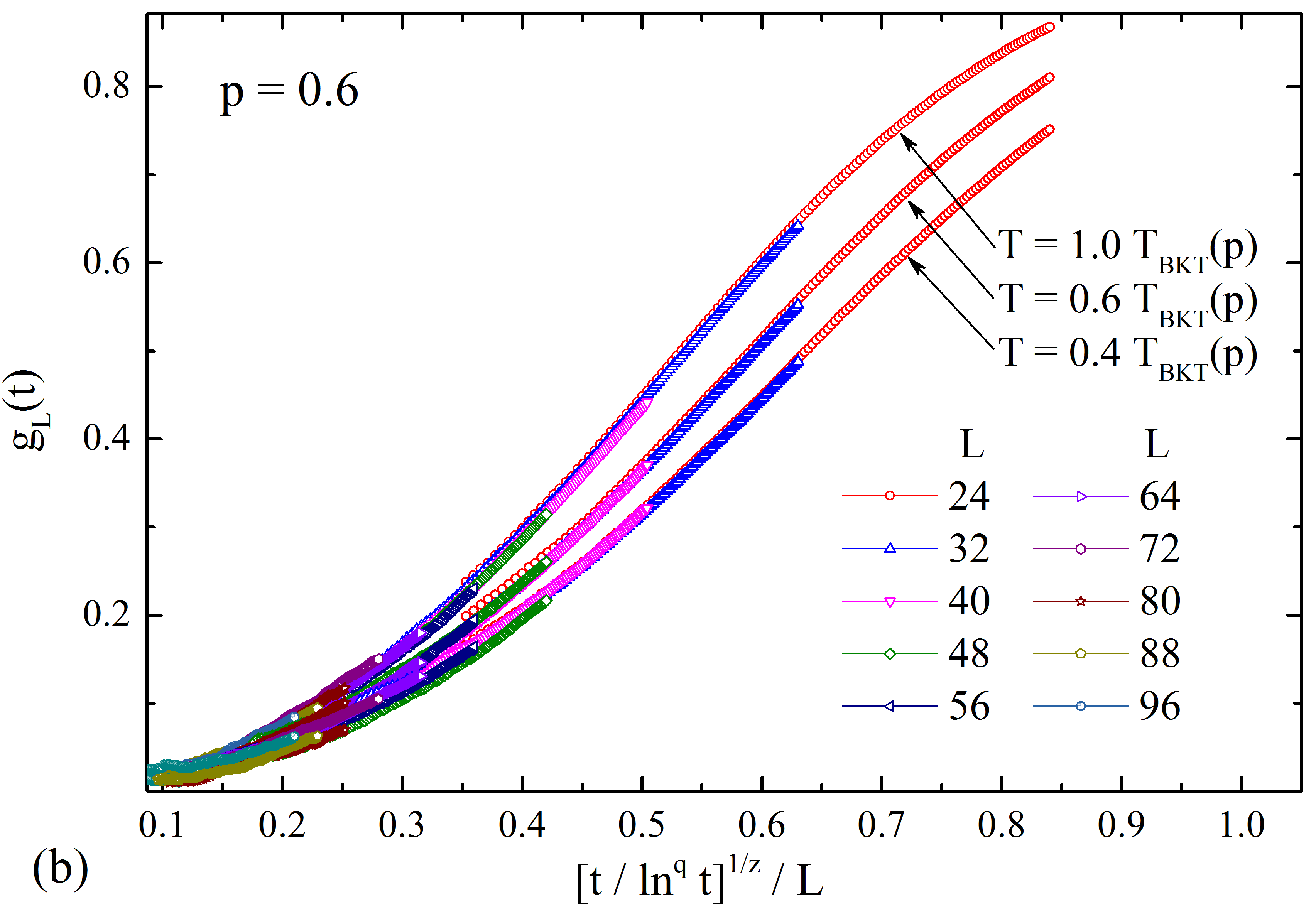}
	\caption{\label{dyn_scale}
		Dynamic scaling dependencies of time-dependent Binder cumulant $g_{L}(t)$
		for system with spin concentrations $p = 0.9$ (a) and $0.6$ (b).
		Temperatures $T$ $=$ $\TBKTp$, $0.6\:\TBKTp$ and $0.4\:\TBKTp$.
		Model dependence $\xi_{\rm m}(t) \sim (t / \ln^{q} t)^{1/z}$,
		where $q = p^{3/2}$ and $z = 2$.
		Results for spin concentrations $p = 0.8$ and $0.7$ see in {\it Supplementary materials}.
	}
\end{figure*}

A well-known relation $\rho_{\rm s}(\TBKT) = 2\TBKT/\pi$ \cite{Korshunov,Kosterlitz_Thouless} 
is satisfied only for the small defect concentrations $p$ $=$ $0.9$, $0.8$ and it is not satisfied in the range of statistical error
for large defect concentration $0.7$, $0.6$. These may indicate a complex dependence $\TBKT(L)$ on the linear size $L$.
The negative stiffness was observed in \cite{NegStiff_1,NegStiff_2,NegStiff_3,NegStiff_4}, however
for the two-dimensional XY-model with annealed defects it was revealed for the first time.
In 2003 Pajda, \textit{et al.}~\cite{NegStiff_1} carried out to study the spin-wave stiffness 
in $\rm Fe$, $\rm Co$ and $\rm Ni$ by the \textit{ab initio} calculations and was shown 
that a negative value of the spin stiffness takes place for the bcc $\rm Fe$ at some 
values of the sumation cutoff parameter $R_{\rm max}$ (see FIG. 2 in \cite{NegStiff_1}).
In \cite{NegStiff_2} it was found that Drude weight $D$ (analogue of transverse stiffness for the ground state of the system) 
has negative value for half-filled one-dimensional Hubbard rings of length $4 \times n$ 
and corresponds to a paramagnetic current response 
(see the discussion in \cite{NegStiff_3}).
The investigation of quantum phase transition in antiferromagnets on triangular lattice 
within the Hubbard model \cite{NegStiff_4}
was shown that electron doping creates a negative contribution
to the spin stiffness of system.

The stiffness can be described as a change of
the free energy $F$ by a twist $\phi$ between every pair of neighbouring lattice cites.
The transverse stiffness is described by the second derivative of the twisted
free energy $F$ with respect to the twist 
$\rho_{\rm s} \sim \frac{d^{2} {F(\phi)}}{d{\phi}^{2}}$,
where $F(\phi)$ is the free energy of the twisted Hamiltonian.
The case of negative stiffness $\rho_{\rm s} < 0$ corresponds to the situation, when the system is unstable with respect to the twist.

The temperature dependence of defects cluster sizes (FIG.~\ref{CS_eq})
shows that defects in the equilibrium state are distributed on a lattice nonuniformly.
There are no free vortices in the thermodynamic equilibrium state.
The equilibrium dynamics in the low-temperature phase $T < \TBKTp$ 
has essentially spin-wave nature~\cite{Berezinskii_book,Korshunov}.
Spin waves in the two-dimensional XY-model represent the long-wave part of the vortex 
excitations spectrum when they are connected into vortex pairs~\cite{Berezinskii_book}.
The process of annealing of defects is conected with the presence an attractive effective potential between defects.
The carried out calculations of equilibrium values of $S_{\rm m}(p,T)$ and $S_{\rm av}(p,T)$ don't show abnormal temperature dependence both.
However, the equilibrium values of sizes of the largest clusters $S_{\rm m}(p,T)$ 
satisfied to scaling behavior 
\begin{equation}
	S_{\rm m}(p,T)|_{L} = \left.S_{\rm m}(p,T)\right|_{L \rightarrow \infty} - \delta(p,T) L^{-\sigma(p)},
\end{equation}
where $L$ is linear size of lattice. The best collapse of data was achieved with values of exponent $\sigma(p) = 0.1 \times (p - 1/2)$.
The results of the calculation $S_{\rm m}(p,T) |_{L \rightarrow \infty}$ and
$L^{-\sigma(p,T)}$ are presented in FIG.~\ref{CS_eq_scale}.
The best collapse for the dependence $\delta(p,T)$ vs $S_{\rm m}(p,T) |_{L \rightarrow \infty}$ are shown in FIG.~\ref{CS_eq_scale}(b).
Should be noted that $\lim_{L \rightarrow \infty} S_{\rm m}(p,T) \neq \infty$.


The investigation of non-equilibrium vortex annealing of structural disorder in the two-dimensional XY-model
in this work is focused on the dynamic dependencies of size of the largest defects cluster $S_{\rm m}(t)$
and the averaged size of defects clusters $S_{\rm av}(t)$ for the different spin concentrations $p$. 
The high-temperature initial state $T_{0} \gg \TBKTp$ was used and the dynamics of the non-equilibrium critical relaxation was mainly vortical,
so the initial non-equilibrium concentration of uncoupled vortices is much greater than for equilibrium state for low-temperature phase 
when concentration has value near zero. 
In FIG.~\ref{CS_noneq_p09} the results for $p = 0.9$ are shown.
Results for the other spin concentrations are presented in {\it Supplementary materials}. 
It doesn't demonstrates a principle differences with case $p = 0.9$ but slowing down of relaxation is appeared with increasing 
the defect concentration $c = 1 - p$ only.

The results show that the subsystem of the large clusters relax more slowly
then whole subsystem of clusters. The dynamic dependencies $S_{\rm m}(t)$ reach the plateau
an order of magnitude longer then $S_{\rm av}(t)$. This difference increases with decreasing of temperature
and the ``inertial" properties of growth start to manifest strongly especially for the averaged size of defects clusters $S_{\rm av}(t)$. Thus one can conclude that the nature of defects cluster growth process is determined by the coarsening.
The ``inertial" properties of clusters growth disappeared at $\TBKTp$ point.

It should be noted that $S_{\rm av}(t)$ is independent of linear size $L$ actually. 
However, $S_{\rm m}(t)$ demonstrates a strong dependence of $L$ (see FIG.~\ref{CS_noneq_p09}).


A dynamical scaling phenomena is essentiall part of non-equilibrium critical relaxation.
We study the presence of dynamical scaling in time-dependent behavior of a cumulant
\begin{equation}
	g_{L}(t) = 2 - \frac{[\langle m^{4} \rangle]}{[\langle m^{2} \rangle]^{2}},
\end{equation}
where $m$ is the magnetization.
We show that the collapse of dynamic data was revealed for $g_{L}(t)$ vs $(t / \ln^{q} (t/t_{0}))^{1/z}/L$ dependencies (see FIG.~\ref{dyn_scale} and {\it Supplementary materials}), where $q = p^{3/2}$ is new exponent which characterized influence of defects. 
For the pure system $p = 1.0$ dynamical scaling corresponds to the results of Bray et al.~\cite{Bray}. 
It was shown~\cite{Bray} that the dynamic dependence $\xi \sim (t / \ln (t/t_{0}))^{1/2}$ 
arises from the solution of equation $d\xi/dt \sim \rho_{\rm s} \Gamma / [\xi \ln (\xi/a)]$.
This solution arise in case $t \gg a^{2}/(4 \rho_{\rm s} \Gamma)$,    
where $W(x) \simeq \ln x - \ln\ln x$, $W(x)$ -- Lambert $W$-function.
The obtained dynamic dependence $(t / \ln^{q} (t/t_{0}))^{1/2}$ 
can be explained by modification of an expression for 
the friction constant $\gamma(R) = (\pi/\Gamma) \ln^{q}(R/a)$, where $R$ is core-core distance in vortex pair.
The solution of equation $d\xi/dt \sim \rho_{\rm s} \Gamma / [\xi \ln^{q} (\xi/a)]$
 can be obtained in $\xi \sim (t / \ln^{q} (t/t_{0}))^{1/2}$ with assumption $t \gg a^{2}/(\rho_{\rm s} \Gamma)$.
 



To conclude, we note that the presence of the annealed disorder leads to significant changes in vortices dynamics of the two-dimensional XY-model.
We revealed that there are temperature ranges with negative transverse stiffness for $T > \TBKTp$.
It should be noted that the annealed defects has non-uniform distribution on lattice in the equilibrium state. The annealing process can be described by the attractive potential arises between defects. This attraction has spin wave nature and  it can be observed in the 
equilibrium critical behavior. The process of non-equilibrium annealing is 
accompanied by the coarsening and defects clustering.
These clusters are formed by the cores of vortices.
The dynamic scaling form $(t / \ln^{q} (t/t_{0}))^{1/z}$ for disordered system can be described by change of the friction constant $\gamma(R) = (\pi/\Gamma) \ln^{p^{3/2}}(R/a)$. 
It was shown that the formation of spatial non-equilibrium structures occurs in forms strips and clumps with different dynamic scales.

\acknowledgments
The reported study was supported by RFBR according to the research projects 
18-32-00814, 17-02-00279, 18-42-550003 and grants MK-4349.2018.2, MD-6868.2018.2 
of the Council of the President of the Russian Federation for State Support of Young Scientists. The simulations were supported in through computational
resources provided by the Shared Facility Center
``Data Center of FEB RAS" (Khabarovsk), Supercomputing Center of Lomonosov Moscow
State University, Moscow Joint Supercomputer Center of RAS.

We would like to thank Prof.~Katanin~A.A. for his important remarks and comments.
Also, we would like to thank our colleague Pospelov~E.A. for his help in improving of the 
manuscript, as well as other staff of the our department for various discussions.

\end{document}


\begin{center}
	\textbf{
		Supplementary materials to the paper:\\
		Non-equilibrium vortex annealing of structural disorder\\
		in Berezinskii-Kosterlitz-Thouless dynamics of two-dimensional XY-model
	}
\end{center}	
	
\begin{center}
	\textbf{Transverse stiffness $\rho_{\rm s}(T)$ of pure two-dimensional XY-model}.
\end{center}

\begin{wrapfigure}[20]{l}{0.56\textwidth}
	\centering
	\includegraphics[width=0.52\textwidth]{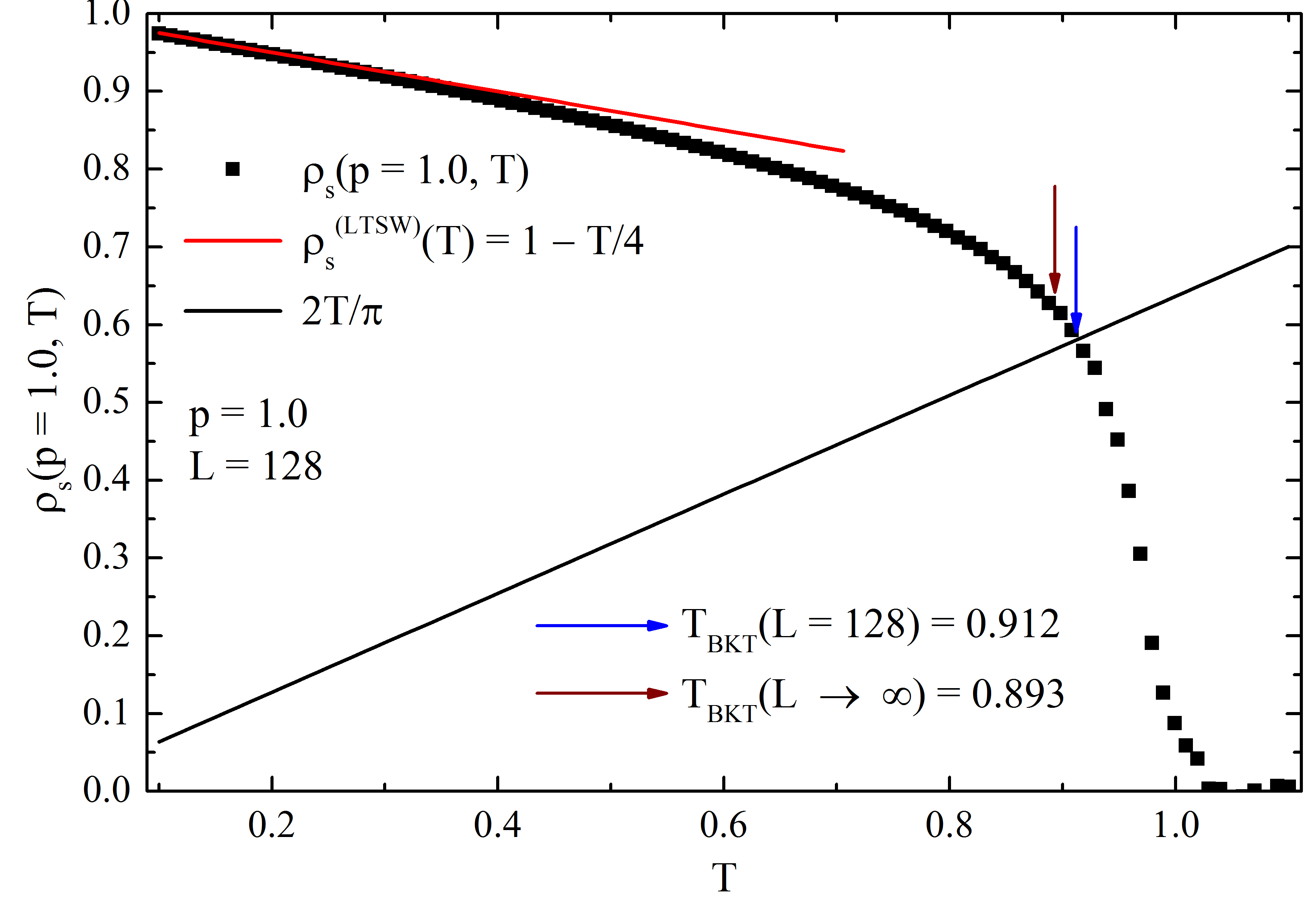}
	\caption{\label{rho_s_verification}
		The temperature dependence of transverse stiffness $\rho_{\rm s}(p,T)$
		for pure system. Blue line demonstrates the linear spin-wave 
		approximation $\rho_{\rm s}^{\rm (LTSW)}$.
		These results demonstrate the correctness of the simulation program.
	}
\end{wrapfigure}

Simulation of critical dynamics of the two-dimensional XY-model in this work
was carried out by Metropolis algorithm.
The simulation results for a pure system (see FIG.~\ref{rho_s_verification}) 
show good agreement with the results of other works (see review~\cite{Korshunov}), 
which shows the correctness of the approach used in this work.
In particular, there is a good agreement on the BKT transition temperature $\TBKT$
by zeroing free energy of vortex condition $\rho_{\rm s}(\TBKT) = 2 \TBKT / \pi$ and 
the low-temperature spin-wave approximation $\rho_{\rm s}^{\rm (LTSW)} = 1 - T/4$
at $T \ll \TBKT$ for pure system. This is the low-temperature solution 
of the Pokrovsky-Uimin equation $\rho(T) = {\rm exp}(-T/4\rho(T))$, which 
used to describe spin-wave behavior of the
two-dimensional XY-model~\cite{Pokrovsky_PLA,Pokrovsky_JETP,PPP_JOPCS_2016}.
\newline
\newline

\begin{center}
	\textbf{Coarsening dynamics at various spin concentrations $p$}.
\end{center}

The investigation of non-equilibrium vortex annealing of structural disorder 
in the two-dimensional XY-model
in this work is focused on the dynamic dependences of size of the largest defects cluster $S_{\rm m}(t)$
and the average size of defects clusters $S(t)$ for the different spin concentrations $p$.

\begin{figure*}[h!]
	\begin{minipage}[t]{0.235\textwidth}
		\includegraphics[width=1.0\textwidth]{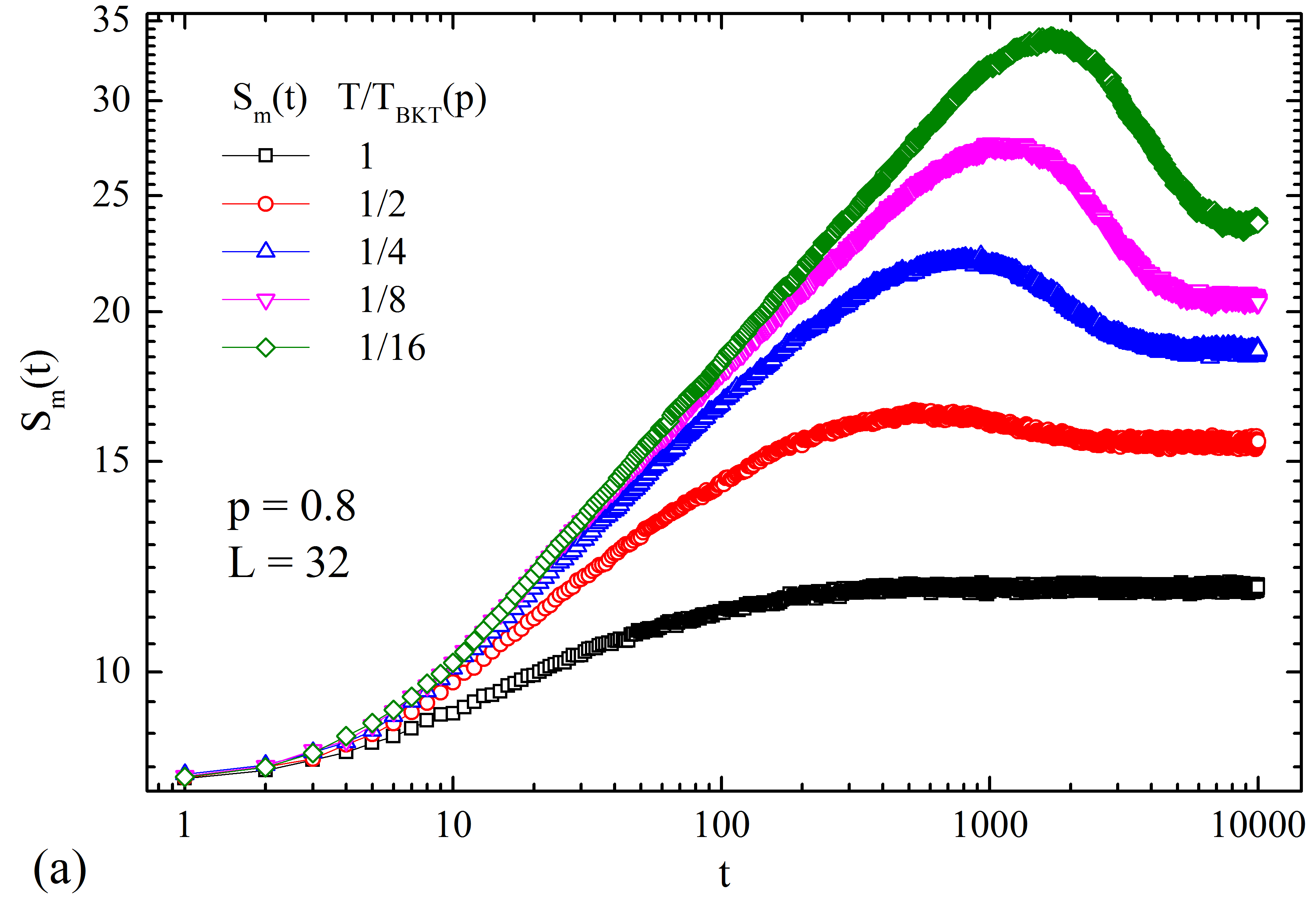}
	\end{minipage}
	\hfill\vspace{-0.1cm}
	\begin{minipage}[t]{0.235\textwidth}
		\includegraphics[width=1.0\textwidth]{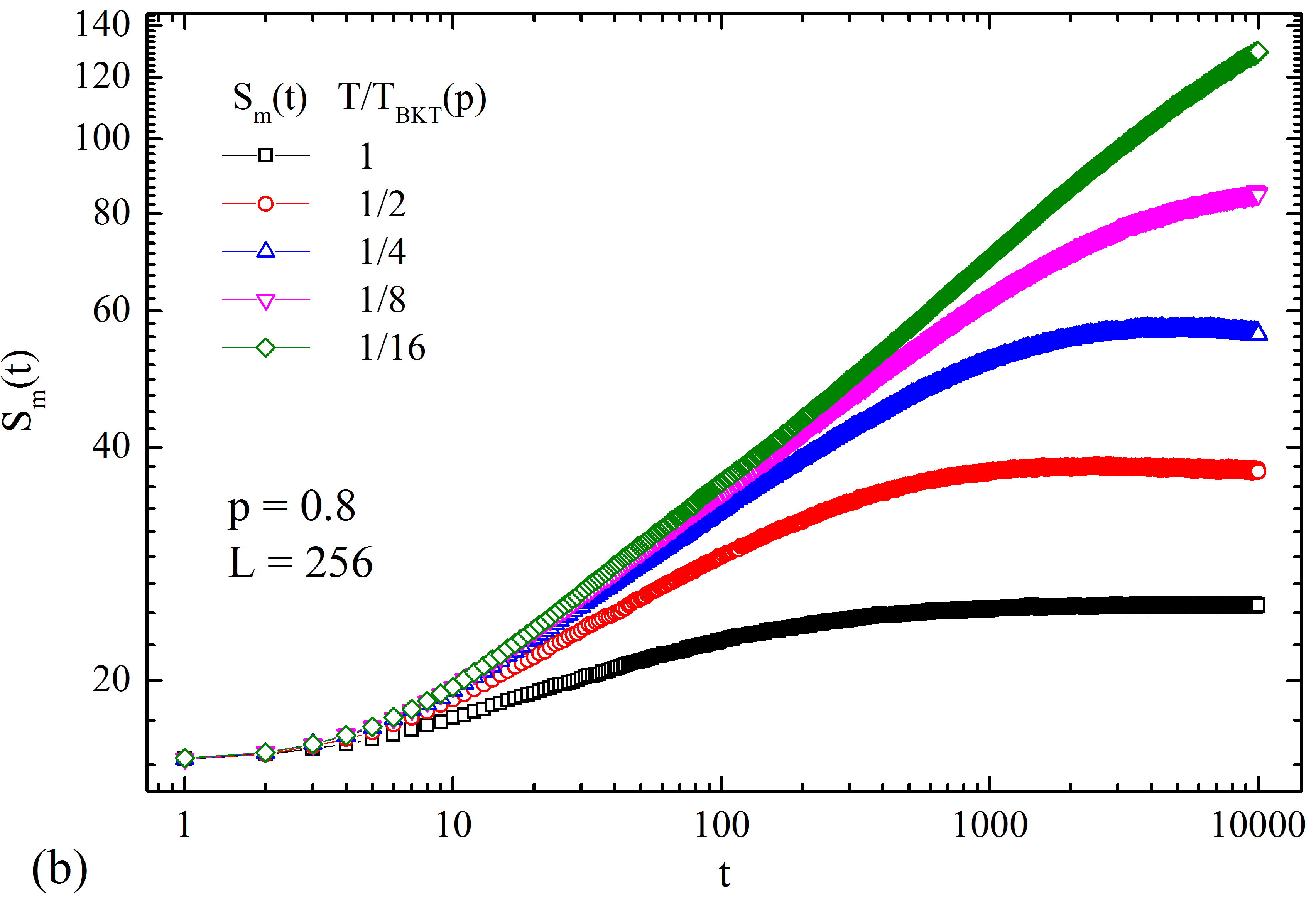}
	\end{minipage}
	\hfill\vspace{-0.1cm}
	\begin{minipage}[t]{0.235\textwidth}
		\includegraphics[width=1.0\textwidth]{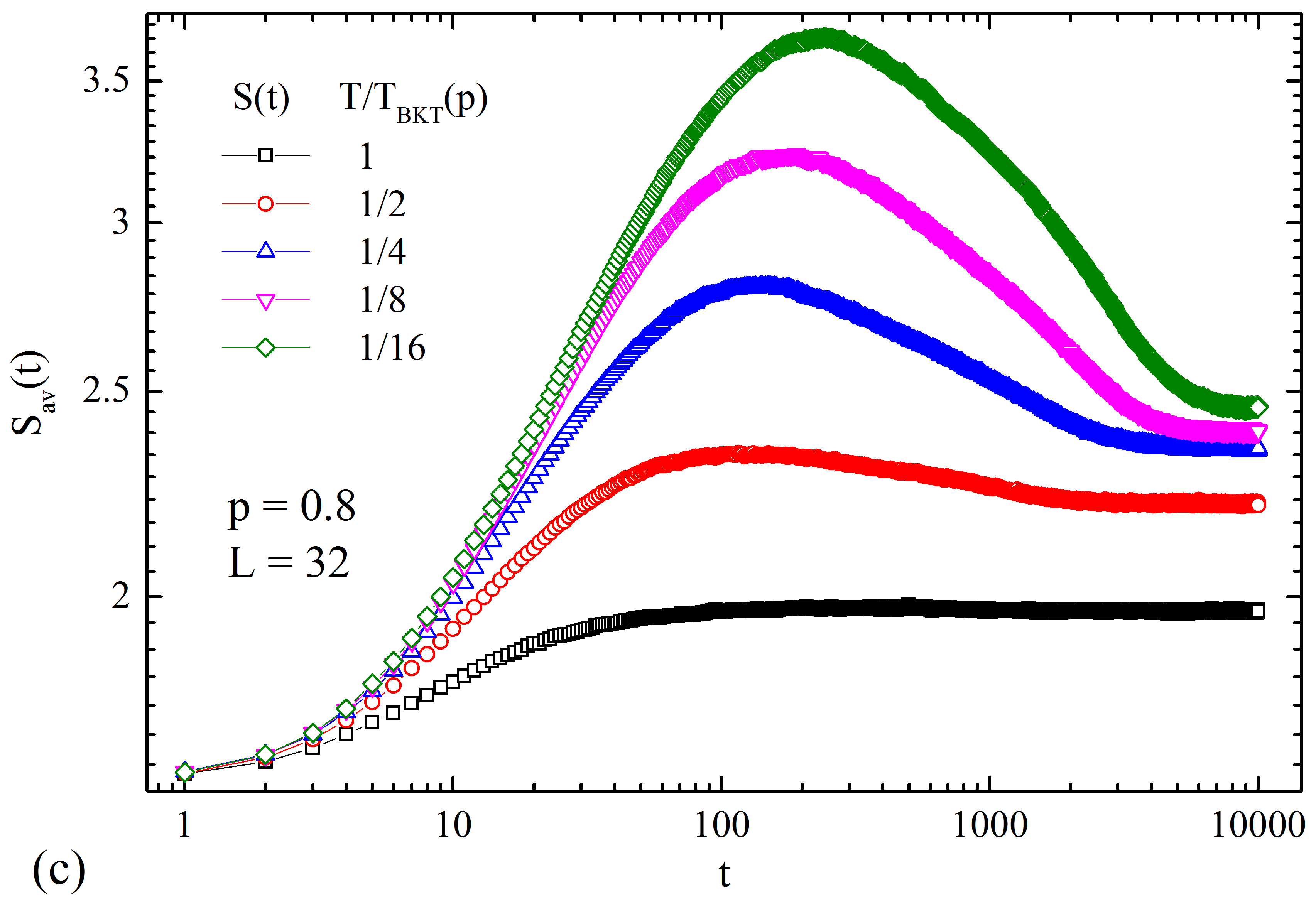}
	\end{minipage}
	\hfill\vspace{-0.1cm}
	\begin{minipage}[t]{0.235\textwidth}
		\includegraphics[width=1.0\textwidth]{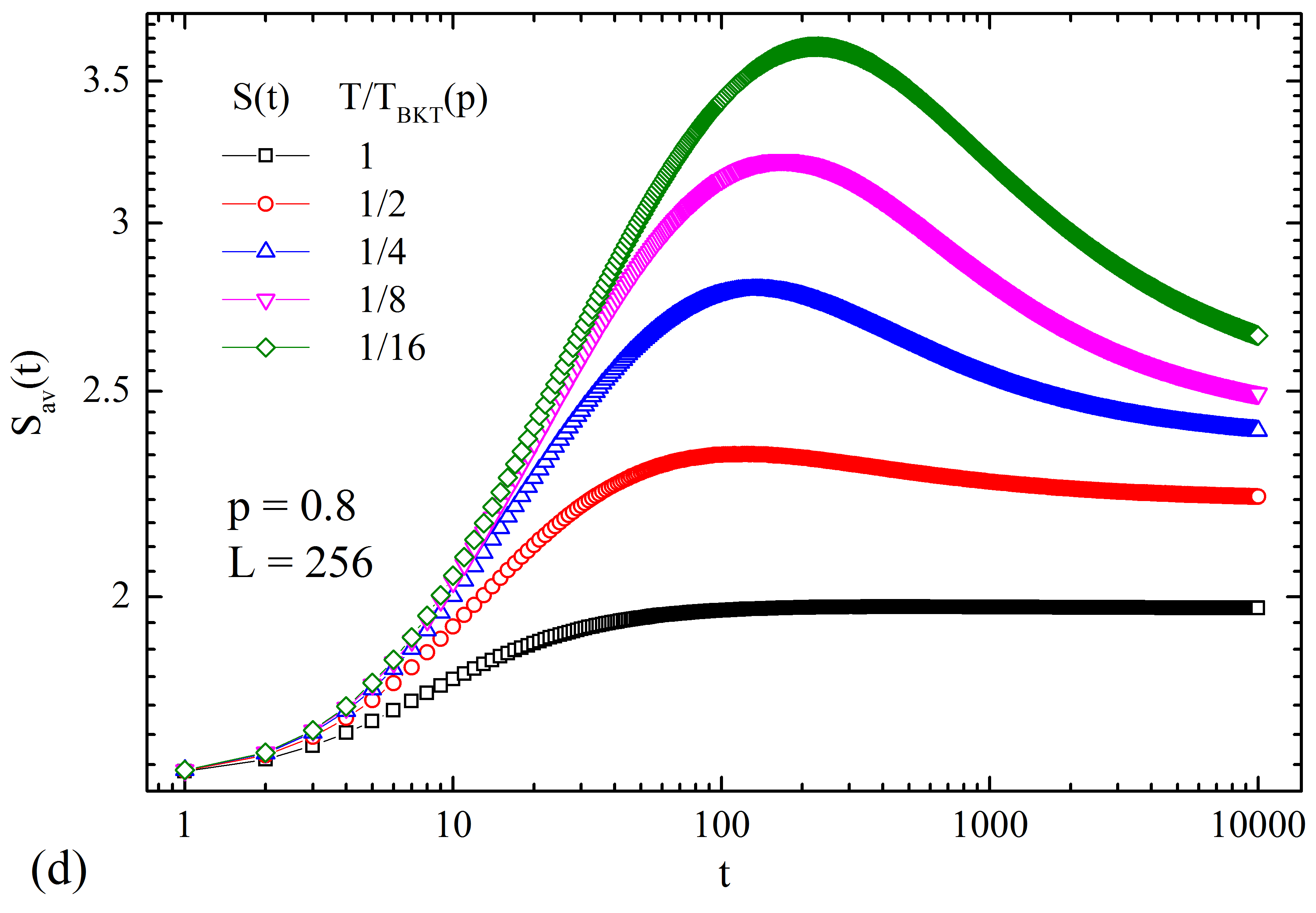}
	\end{minipage}
	\caption{\label{CS_noneq_p08}
		The dynamic dependences of size of the largest cluster of defects $S_{\rm m}(t)$ (a, b)
		and average size of clusters of defects $S(t)$ (c, d) for
		system with spin concentration $p = 0.8$ and linear sizes $L = 32$ (a, c)
		and $L = 256$ (b, d).
	}
\end{figure*}

\begin{figure*}[h!]
	\begin{minipage}[t]{0.235\textwidth}
		\includegraphics[width=1.0\textwidth]{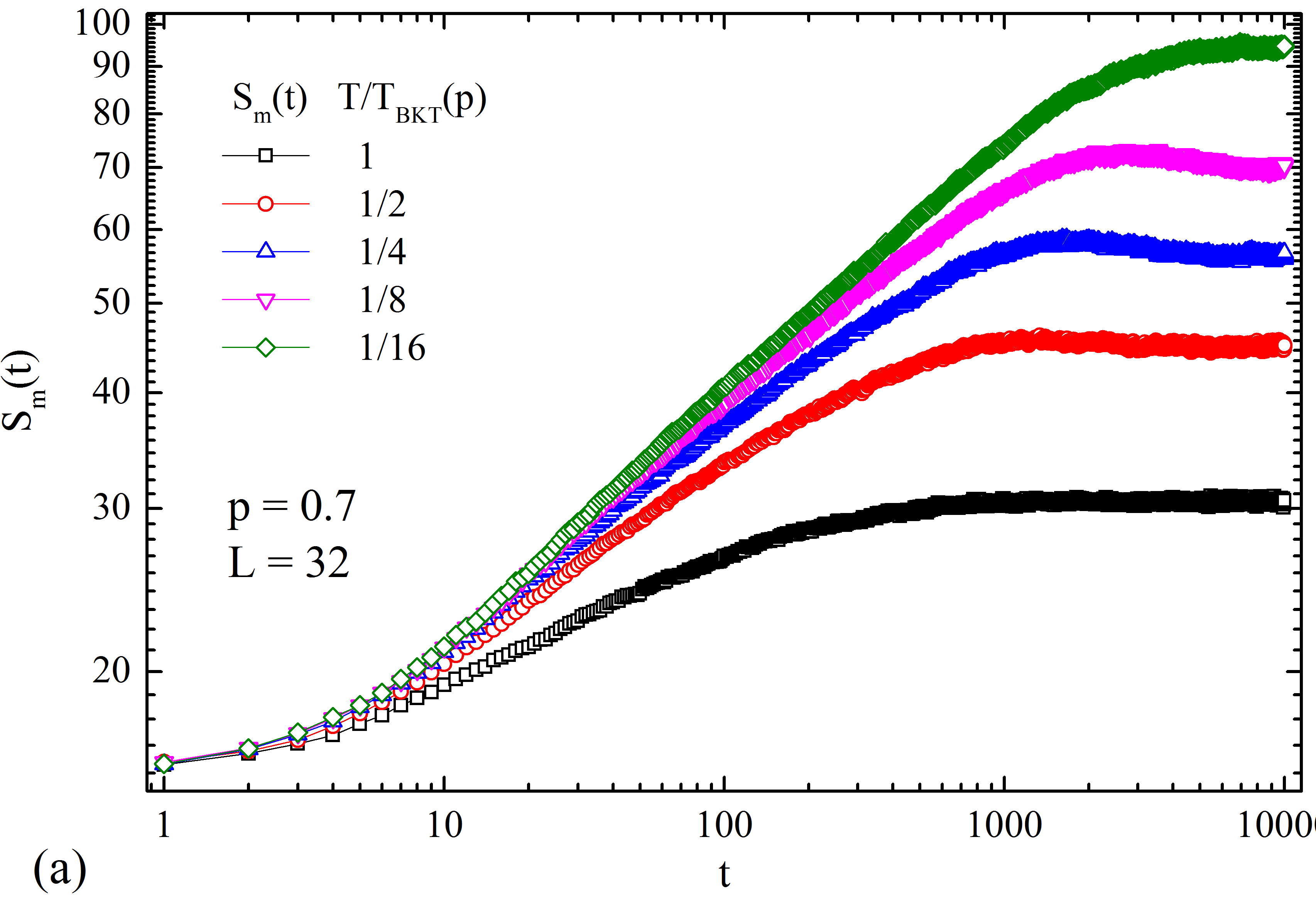}
	\end{minipage}
	\hfill\vspace{-0.1cm}
	\begin{minipage}[t]{0.235\textwidth}
		\includegraphics[width=1.0\textwidth]{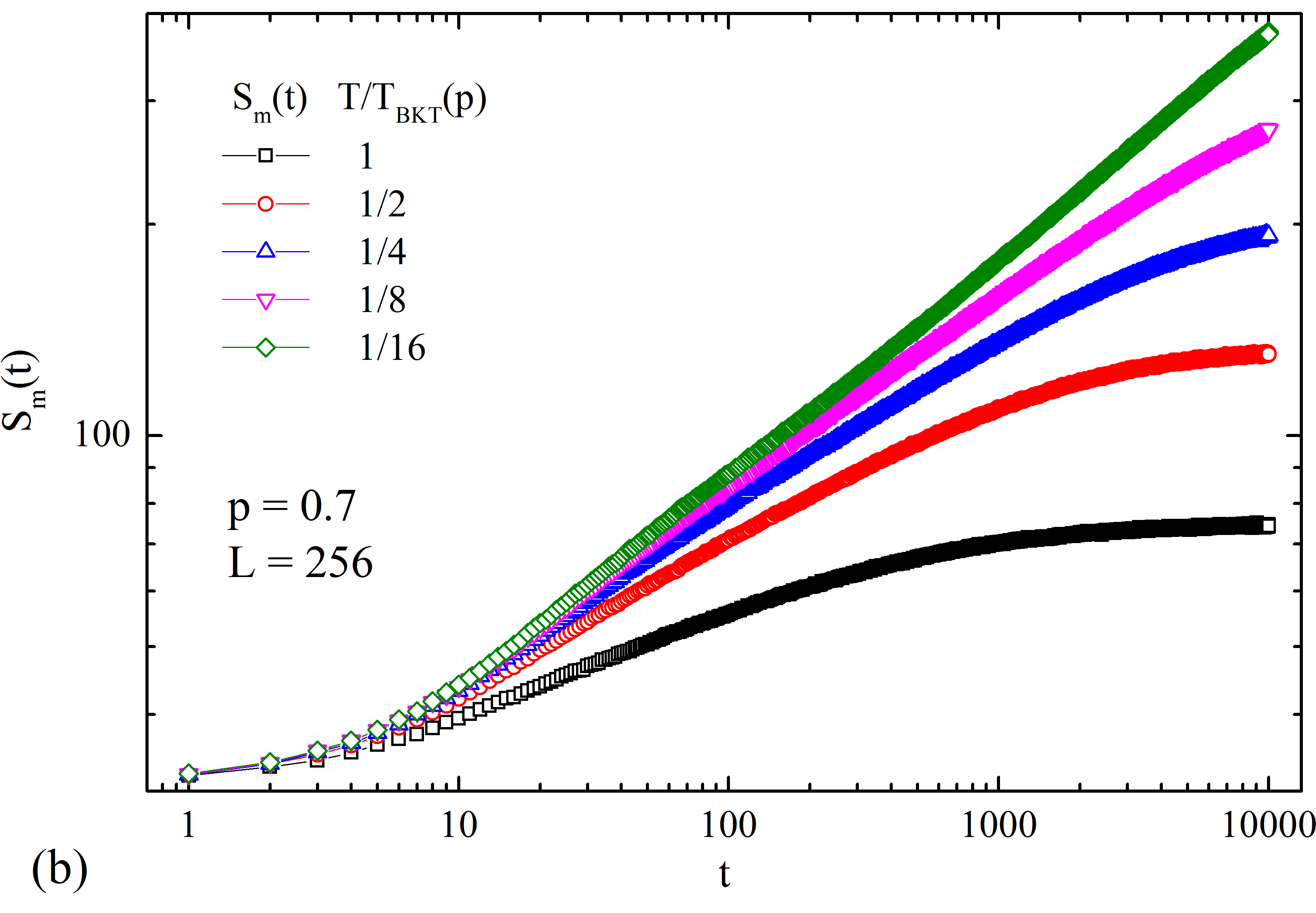}
	\end{minipage}
	\hfill\vspace{-0.1cm}
	\begin{minipage}[t]{0.235\textwidth}
		\includegraphics[width=1.0\textwidth]{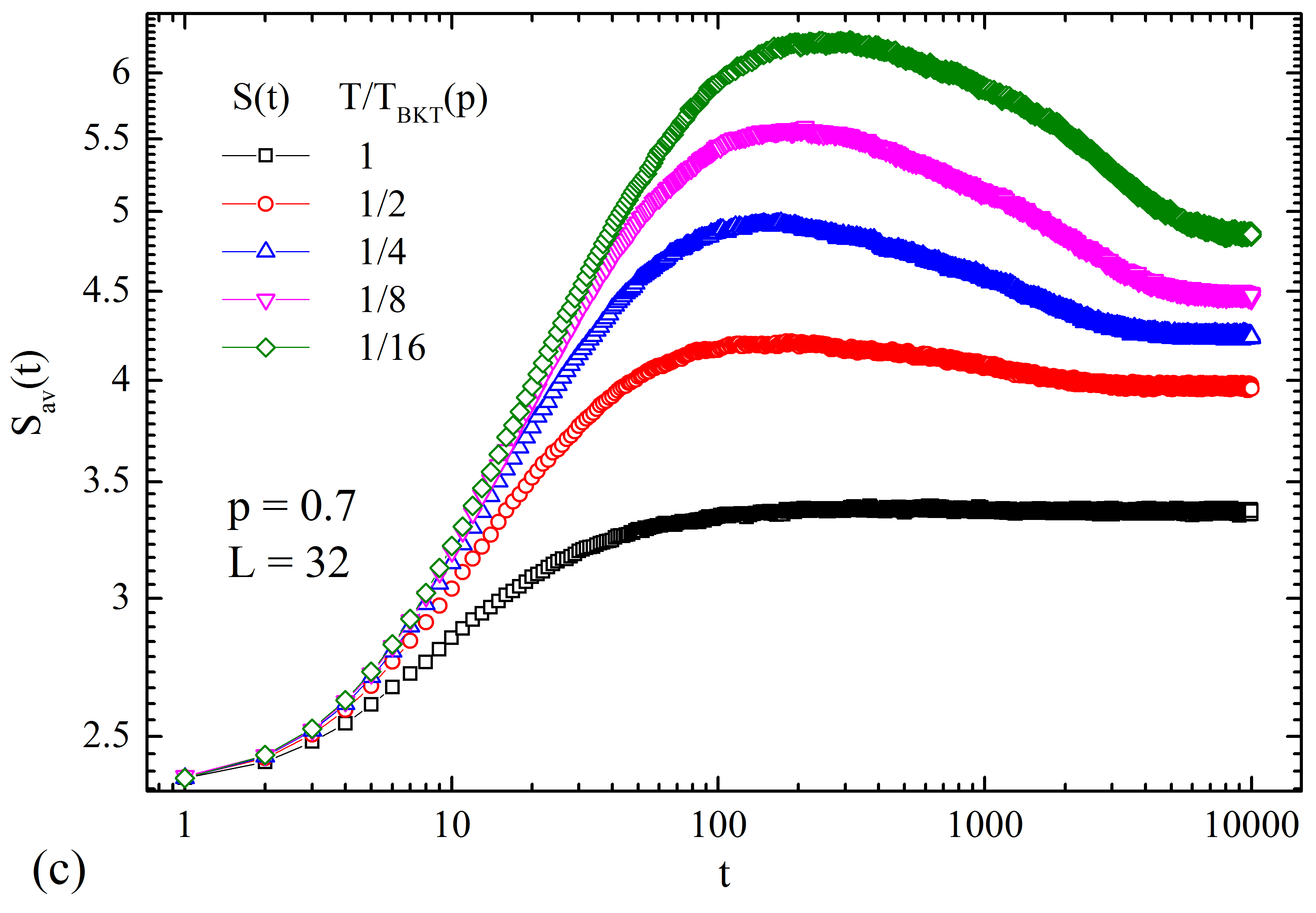}
	\end{minipage}
	\hfill\vspace{-0.1cm}
	\begin{minipage}[t]{0.235\textwidth}
		\includegraphics[width=1.0\textwidth]{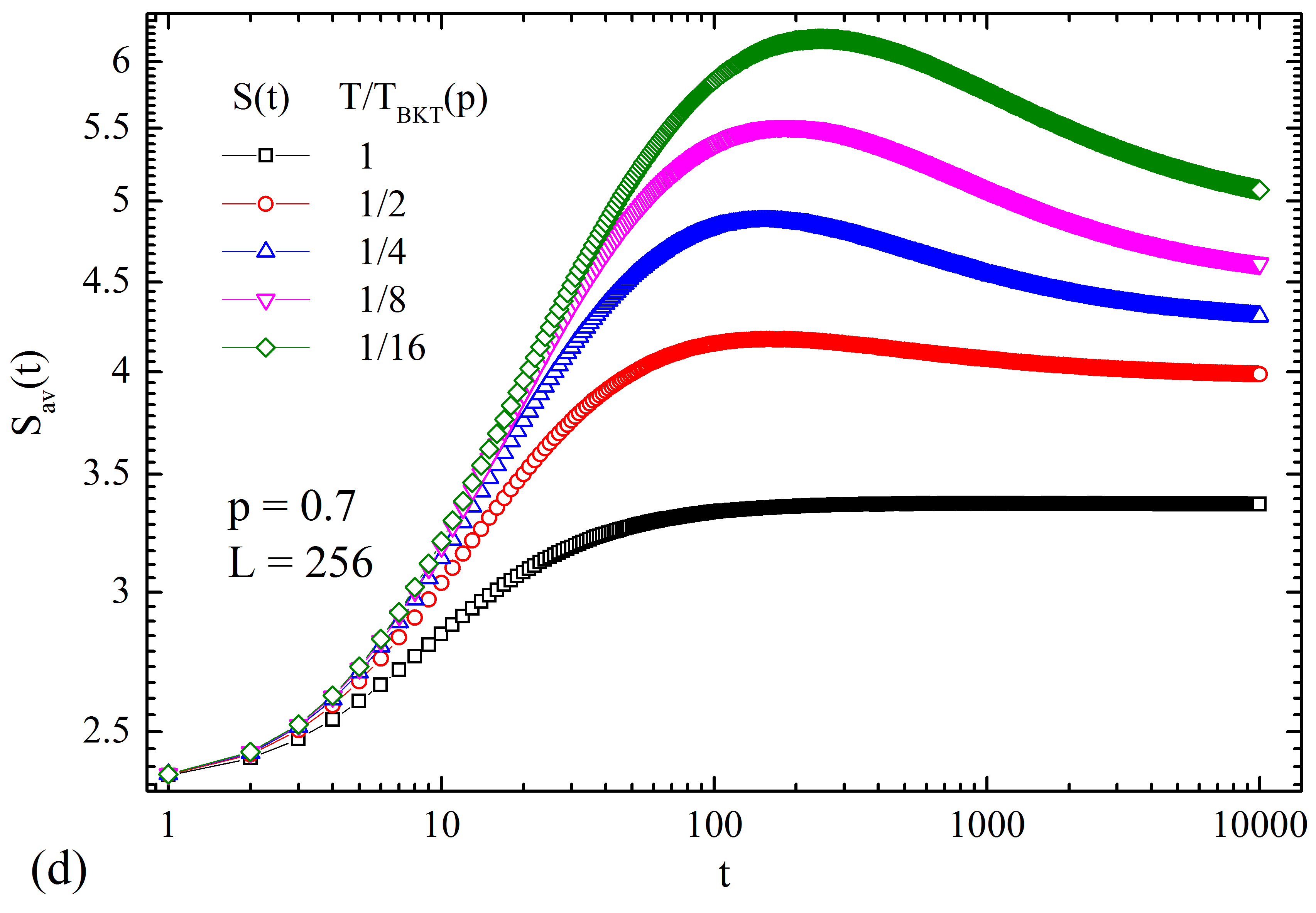}
	\end{minipage}
	\caption{\label{CS_noneq_p07}
		The dynamic dependences of size of the largest cluster of defects $S_{\rm m}(t)$ (a, b)
		and average size of clusters of defects $S(t)$ (c, d) for
		system with spin concentration $p = 0.7$ and linear sizes $L = 32$ (a, c)
		and $L = 256$ (b, d).
	}
\end{figure*}

\begin{figure*}[h!]
	\begin{minipage}[t]{0.235\textwidth}
		\includegraphics[width=1.0\textwidth]{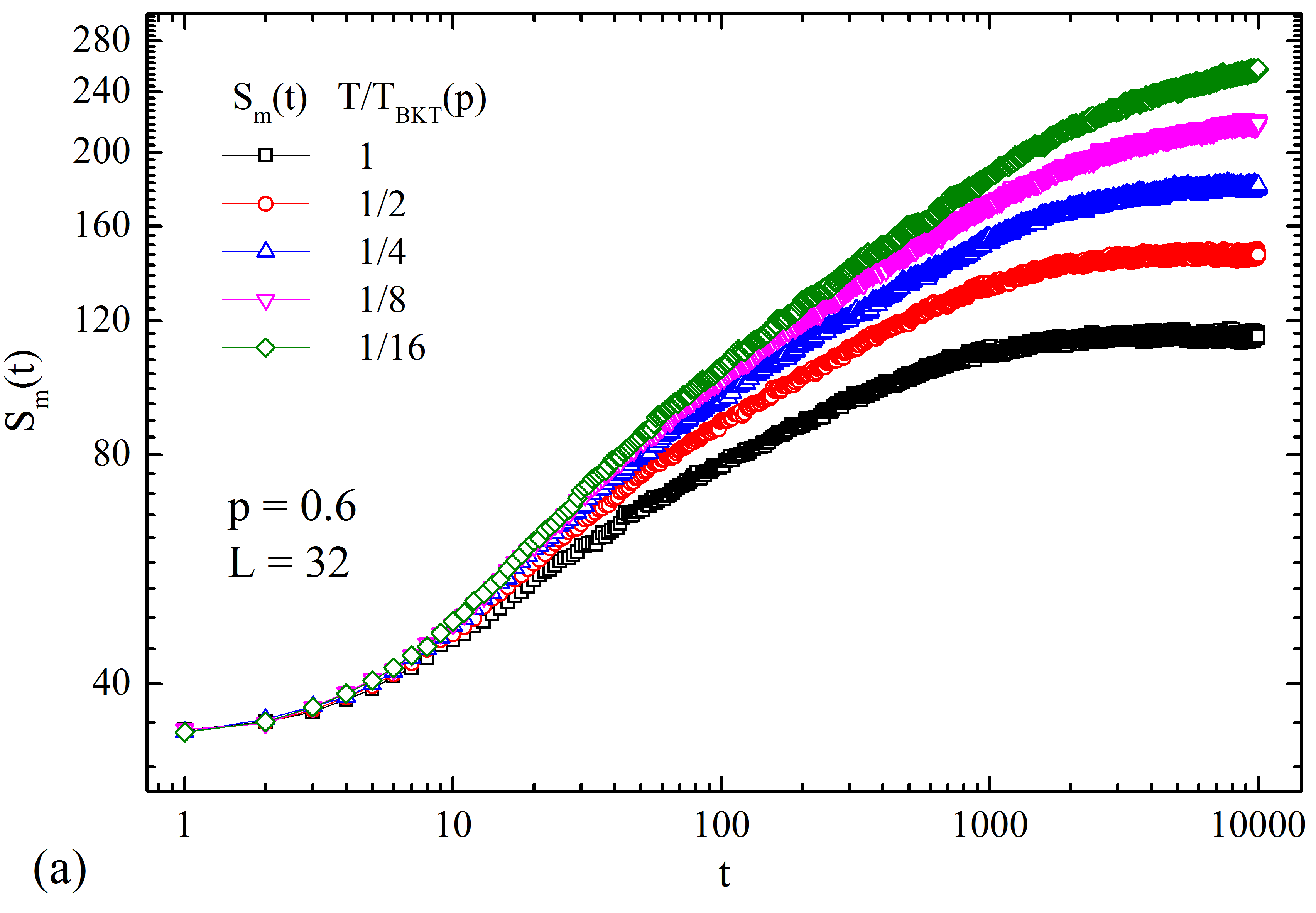}
	\end{minipage}
	\hfill\vspace{-0.1cm}
	\begin{minipage}[t]{0.235\textwidth}
		\includegraphics[width=1.0\textwidth]{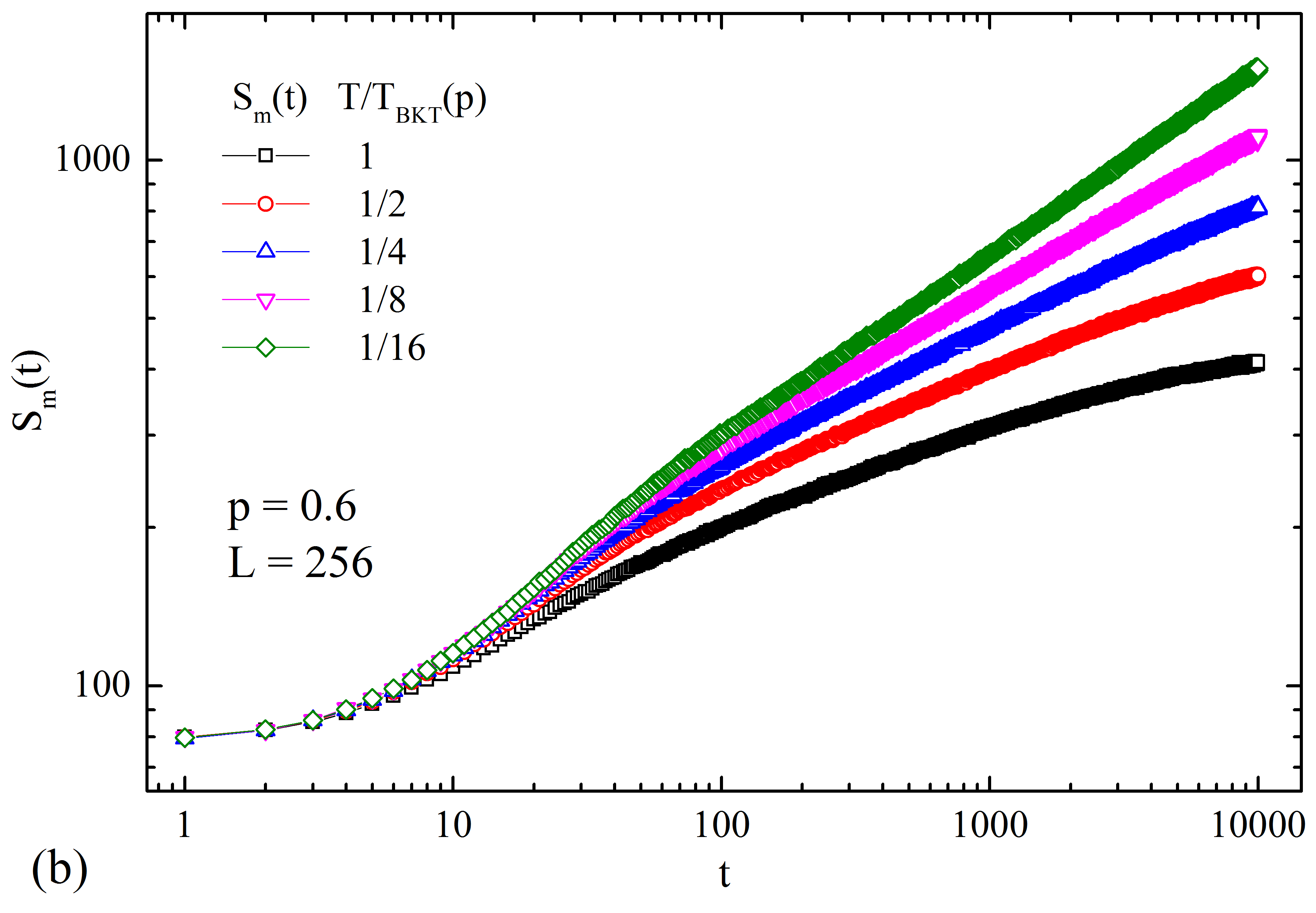}
	\end{minipage}
	\hfill\vspace{-0.1cm}
	\begin{minipage}[t]{0.235\textwidth}
		\includegraphics[width=1.0\textwidth]{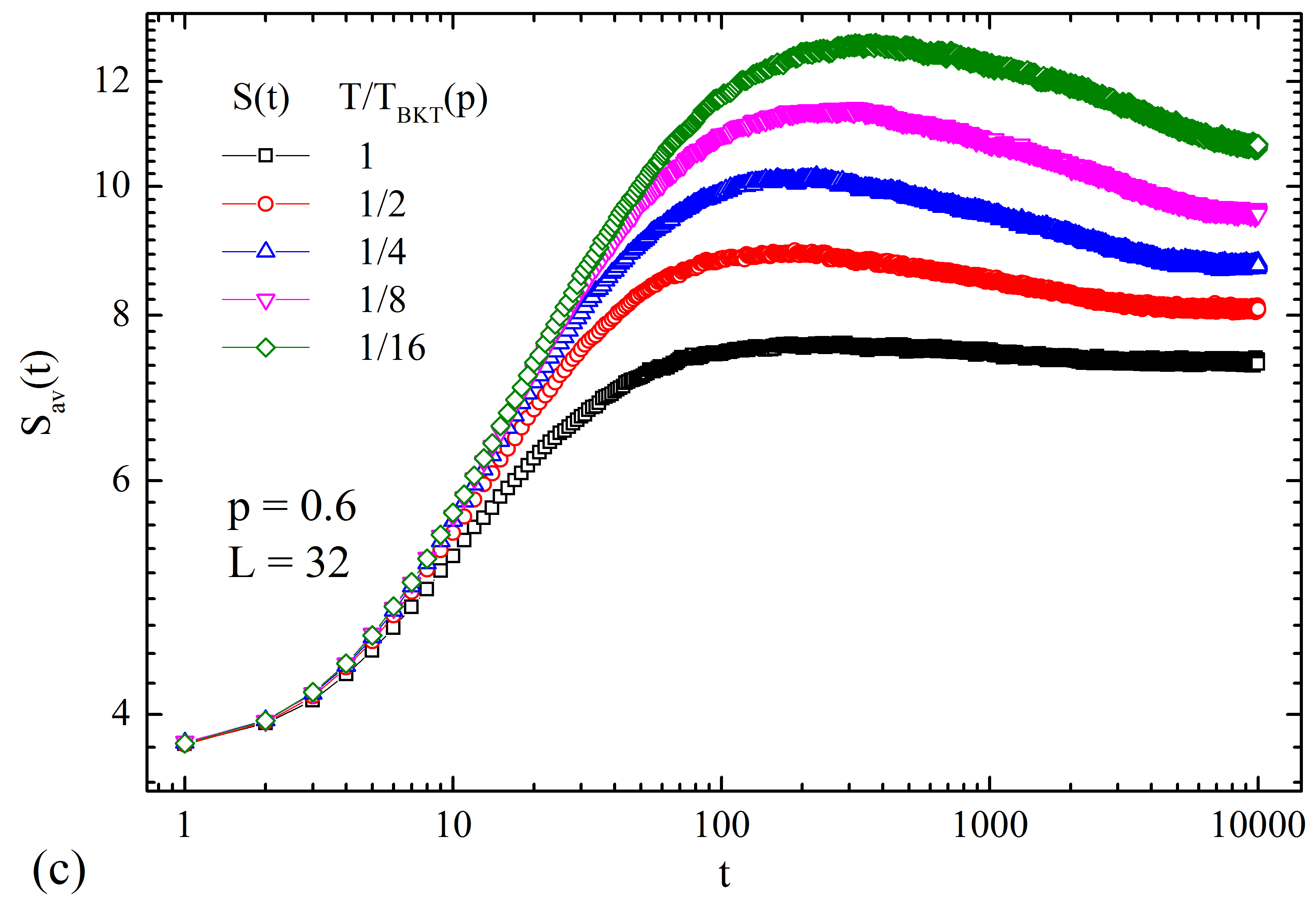}
	\end{minipage}
	\hfill\vspace{-0.1cm}
	\begin{minipage}[t]{0.235\textwidth}
		\includegraphics[width=1.0\textwidth]{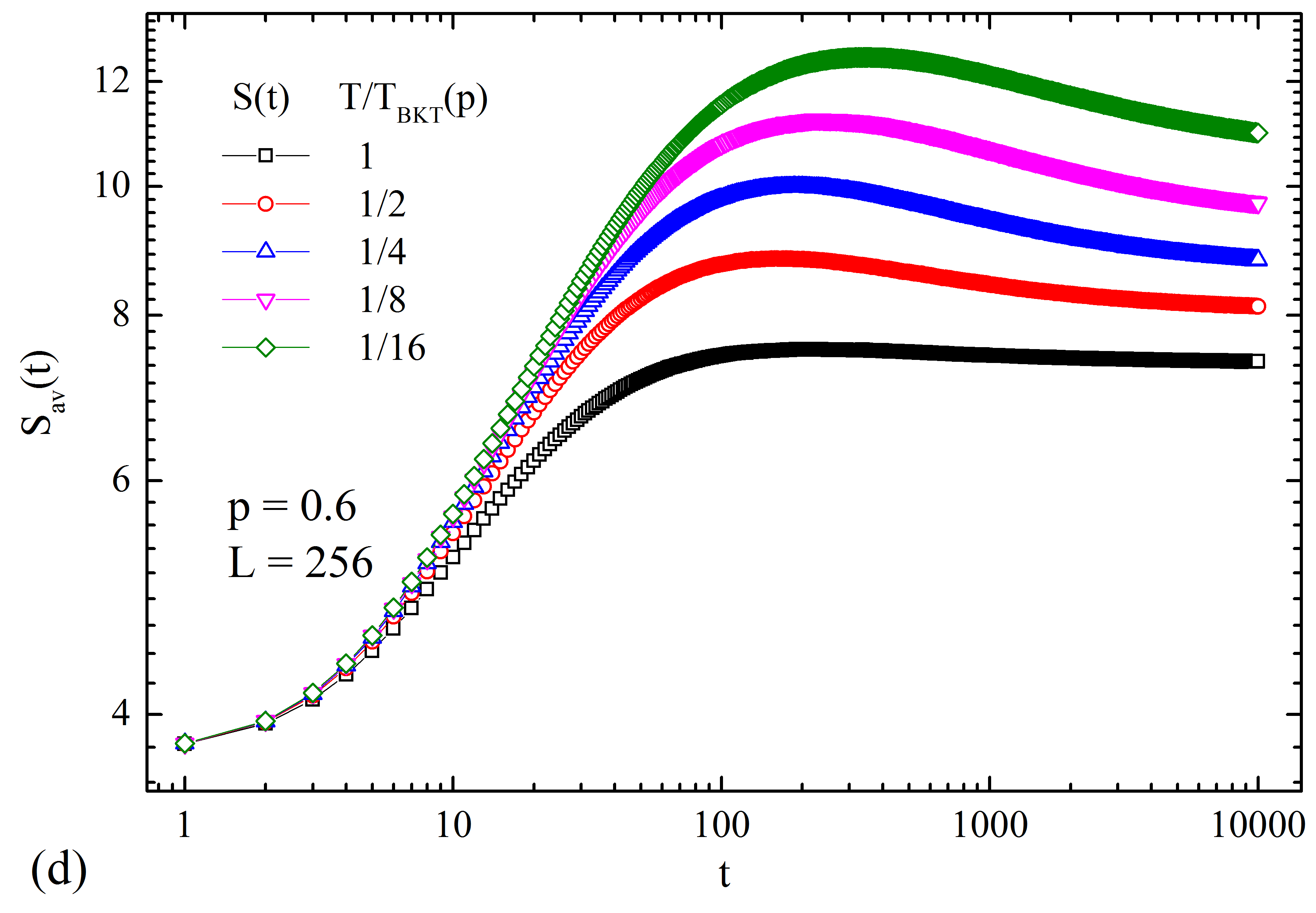}
	\end{minipage}
	\caption{\label{CS_noneq_p06}
		The dynamic dependences of size of the largest cluster of defects $S_{\rm m}(t)$ (a, b)
		and average size of clusters of defects $S(t)$ (c, d) for
		system with spin concentration $p = 0.6$ and linear sizes $L = 32$ (a, c)
		and $L = 256$ (b, d).
	}
\end{figure*}

This appendix presents the simulation results for spin concentrations $p = 0.8$, $0.7$ and $0.6$
(see FIG.~\ref{CS_noneq_p08},~\ref{CS_noneq_p07} and~\ref{CS_noneq_p06}).

The presented results allow us to conclude that the growth 
of largest clusters size $S_{\rm m}(t)$ is slowing down with the defect concentration $c = 1 - p$.
However, it is important to note that the dynamic scale of 
average size of defects clusters $S(t)$ growth 
remains almost unchanged with changes in 
the defect concentration $c$.
At the same time, with increasing defect concentration $c$,
the largest clusters size $S_{\rm m}(t)$ and the average size 
of defects clusters $S(t)$ increase.

The reason for this behavior is not complicated.
The growth of the largest clusters (with size $S_{\rm m}(t)$) 
in the system is due to coarsening of the vortices.
The pinning of defects by vortices occurs as the defect concentration $c$ increases.
When this happens, more defects are clustered in the vortex cores, 
which is accompanied by an increase of $S_{\rm m}(t)$.


\begin{center}
	\textbf{Non-equilibrium scaling properties at various spin concentrations $p$}.
\end{center}

\begin{figure}[ht!]
	\centering
	\begin{minipage}[t]{0.32\textwidth}
	\includegraphics[width=0.98\textwidth]{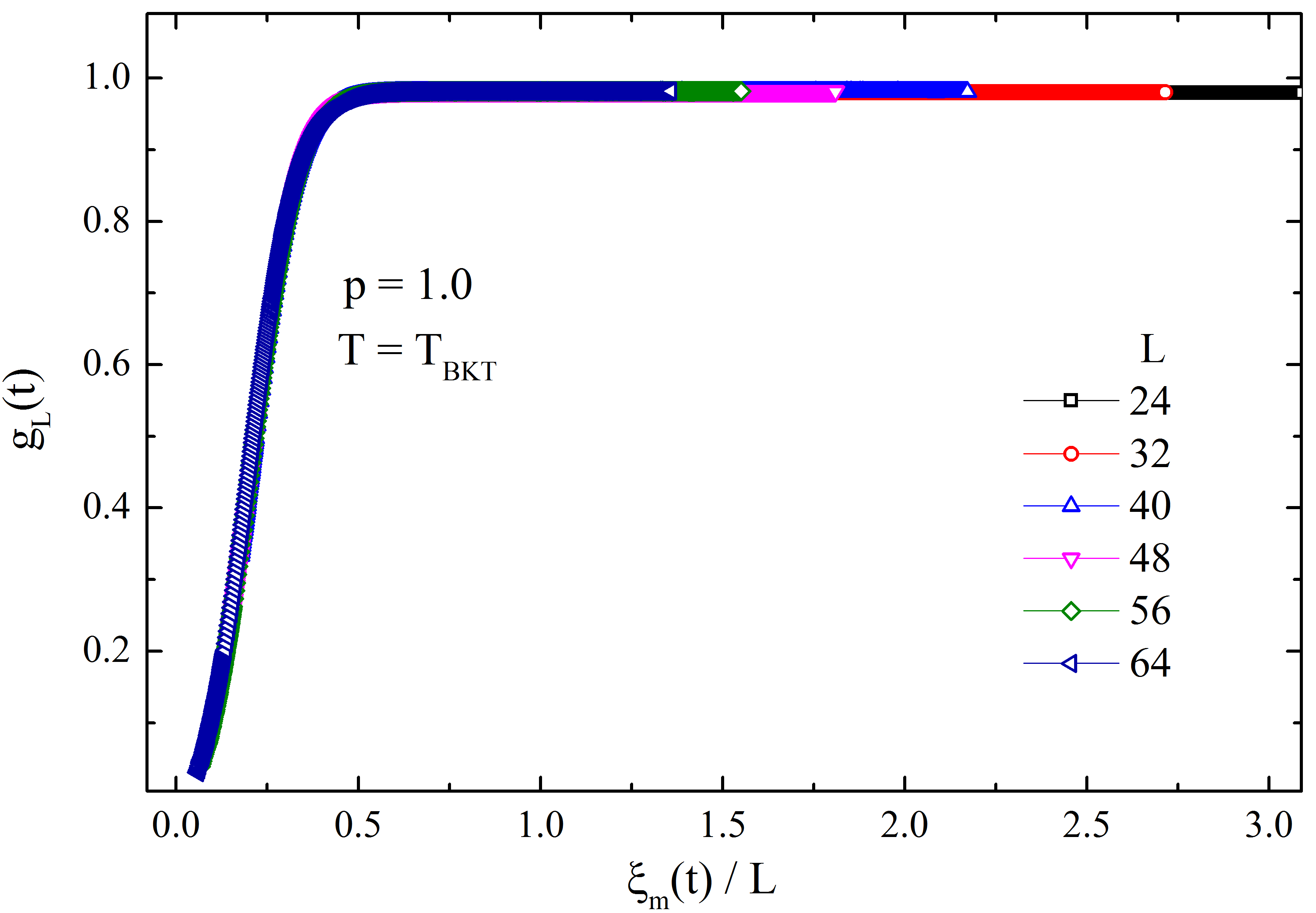}\vspace{-0.1cm}
	\caption{\label{gL_test_BrayData}
		Dynamic scaling dependences of time-dependent Binder cumulant $g_{L}(t)$
		for pure system ($p = 1.0$).
		Model dependence $\xi_{\rm m}(t) \sim (t / \ln t)^{1/z}$, 
		where $z = 2$.
	}
	\end{minipage}\hspace{4.5mm}
	\begin{minipage}[t]{0.64\textwidth}
	\includegraphics[width=0.49\textwidth]{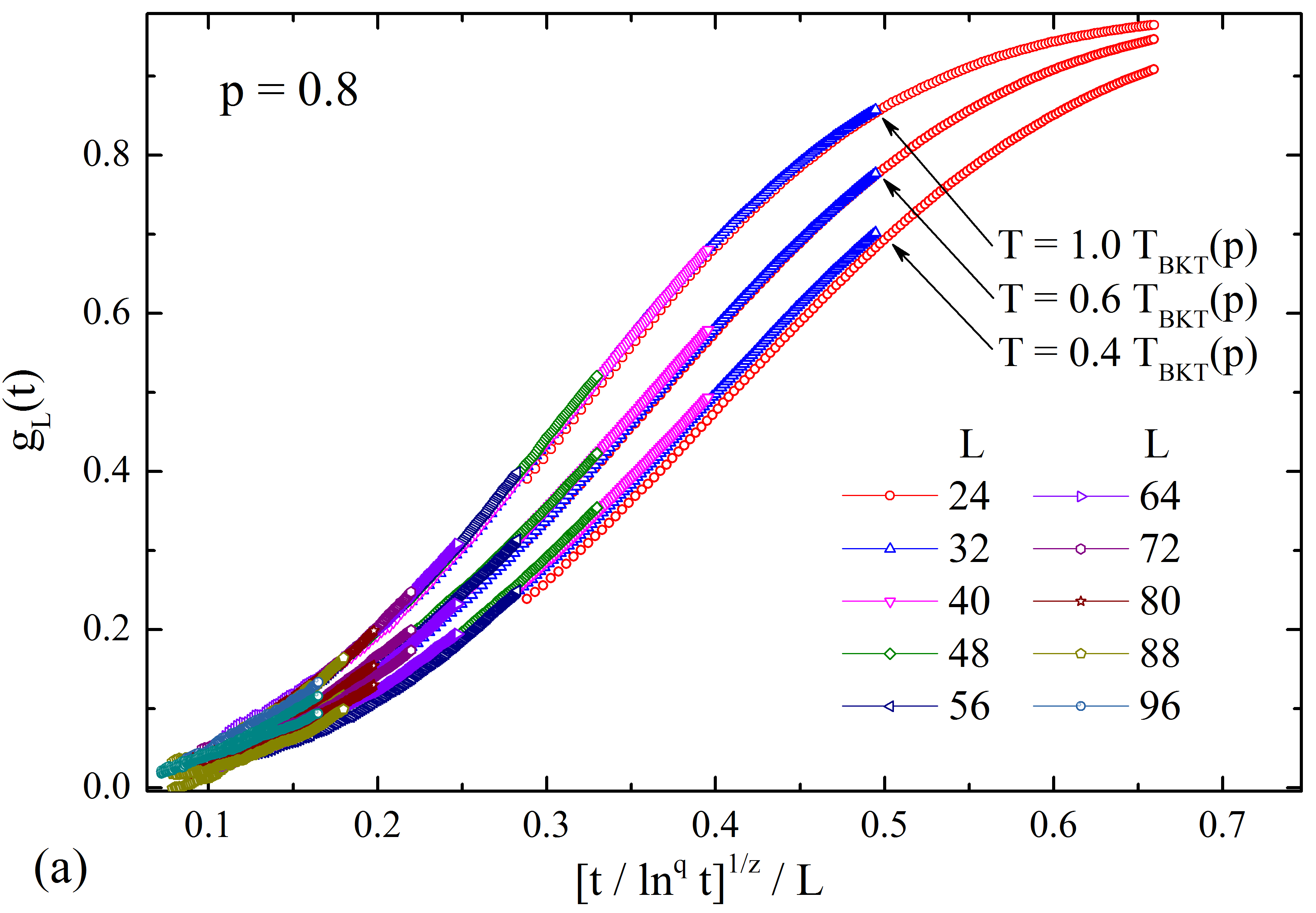}
	\includegraphics[width=0.49\textwidth]{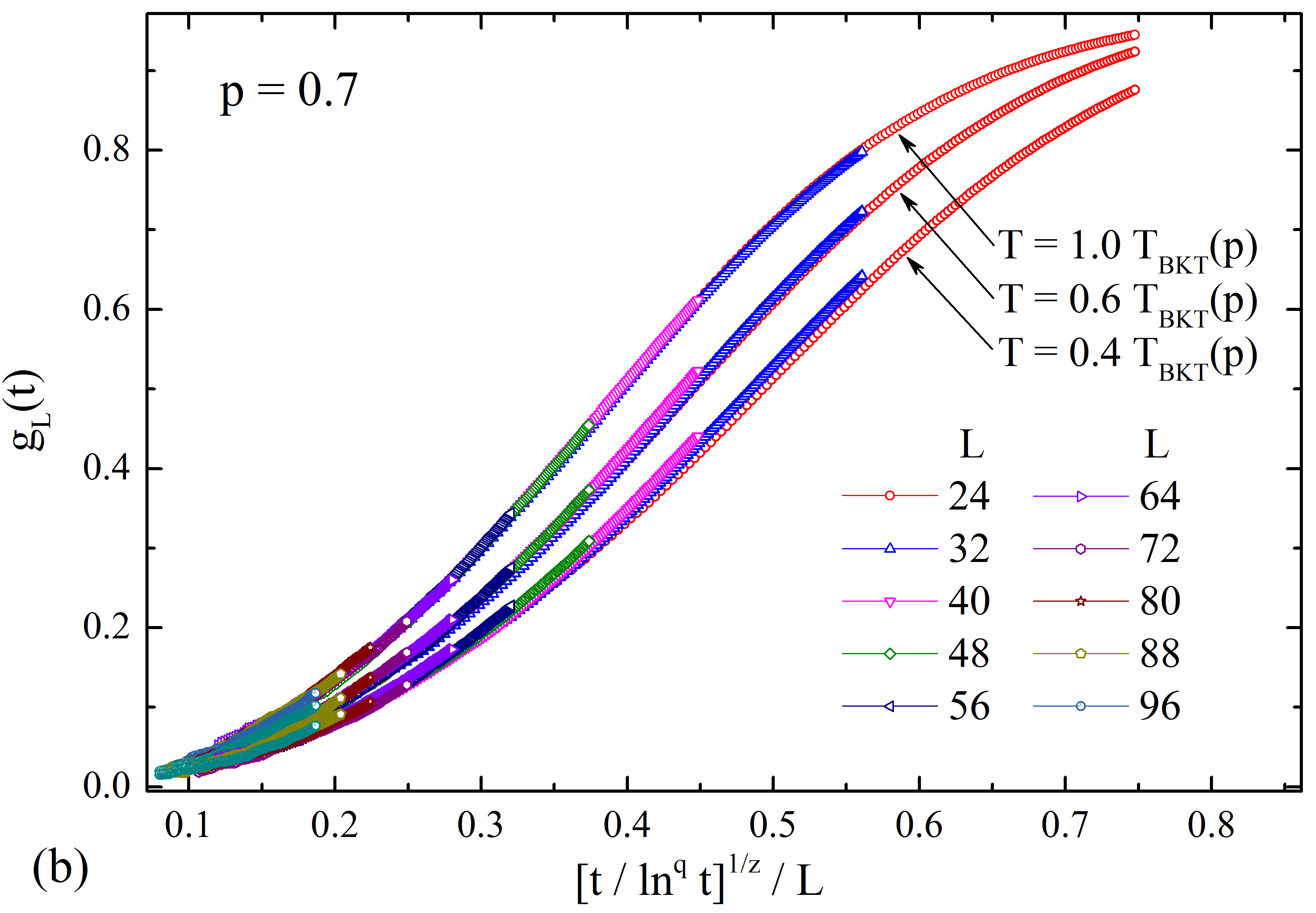}\vspace{-0.1cm}
	\caption{\label{dyn_scale_app}
		Dynamic scaling dependences of time-dependent Binder cumulant $g_{L}(t)$
		for system with spin concentrations $p = 0.8$ (a) and $0.7$ (b).
		Temperatures $T$ $=$ $\TBKTp$, $0.6\:\TBKTp$ and $0.4\:\TBKTp$.
		Model dependence $\xi_{\rm m}(t) \sim \big(t / \ln^{q} t\big)^{1/z}$, 
		where $q = p^{3/2}$ and $z = 2$.
	}
	\end{minipage}
\end{figure}

The simulation results for a pure system (see FIG.~\ref{gL_test_BrayData}) 
show good agreement with the results~\cite{AD_Bray}, 
which shows the correctness of the approach used in this work.

This appendix presents the simulation results of non-equilibrium scaling properties 
for spin concentrations $p = 0.8$ and $0.7$
(see FIG.~\ref{dyn_scale_app}).